\documentclass[useAMS,usenatbib]{mn2e}

\usepackage{graphicx}
\usepackage{lscape}
\usepackage{amsmath}
\usepackage{amssymb}
\usepackage{rotating}

\newcommand\aj{\rmfamily{AJ}}%
\newcommand\araa{\rmfamily{ARA\&A}}%
\newcommand\apj{\rmfamily{ApJ}}%
\newcommand\apjl{\rmfamily{ApJ}}%
\newcommand\apjs{\rmfamily{ApJS}}%
\newcommand\aaps{\rmfamily{A\&AS}}%
\newcommand\aap{\rmfamily{A\&A}}%
\newcommand\mnras{\rmfamily{MNRAS}}%
\newcommand\nat{\rmfamily{Nature}}%
\newcommand\pasp{\rmfamily{PASP}}%

\title[Near-UV extended light in Seyfert galaxies]{On the nature of the near-UV extended light in Seyfert galaxies}
\author[Mu\~noz Mar\'{i}n et al.]{V.~M.~Mu\~noz Mar\'{i}n,$^1$\thanks{E-mail: manuel@iaa.es} T.~Storchi-Bergmann,$^2$ R.~M.~Gonz\'alez Delgado,$^1$ \newauthor H.~R.~Schmitt,$^{3,4}$ P.~F.~Spinelli,$^{2,5}$ E.~P\'erez,$^1$ R.~Cid Fernandes$^6$ \\
$^1$Instituto de Astrof\'{i}sica de Andaluc\'{i}a (IAA-CSIC), Granada, Spain\\
$^2$Instituto de F\'{i}sica, Universidade Federal do Rio Grande do Sul, IF, CP 15051, Porto Alegre 91501-970, RS, Brazil\\
$^3$Naval Research Laboratory, Remote Sensing Division, Code 7215, 4555 Overlook Ave. SW, Washington, DC 20375, US\\
$^4$Interferometrics, Inc, 13454 Sunrise Valley Drive, Suite 240, Herndon, VA 20171, US\\
$^5$Universit\"atssternwarte M\"unchen, D-81679 M\"unchen, Germany\\
$^6$Universidade Federal de Santa Catarina, Florian\'opolis, SC, Brazil}

\begin{document}

\date{}

\pagerange{\pageref{firstpage}--\pageref{lastpage}} \pubyear{2008}

\maketitle

\label{firstpage}

\begin{abstract}
We study the nature of the extended near-UV emission in the inner kiloparsec of a sample of 15 Seyfert galaxies which have both near-UV (F330W) and narrow band [OIII] high resolution Hubble images. 
For the majority of the objects we find a very similar morphology in both bands. From the [OIII] images we construct synthetic images of the nebular continuum plus the emission line contribution expected through the F330W filter, which can be subtracted from the F330W images. We find that the emission of the ionised gas dominates the near-UV extended emission in half of the objects. A further broad band photometric study, in the bands F330W ({\it U}), F547M ({\it V}) and F160W ({\it H}), shows that the remaining emission is dominated by the underlying galactic bulge contribution. We also find a blue component whose nature is not clear in 4 out of 15 objects. This component may be attributed to scattered light from the AGN, to a young stellar population in unresolved star clusters, or to early-disrupted clusters. Star forming regions and/or bright off-nuclear star clusters are observed in 4/15 galaxies of the sample.
\end{abstract}



\section{INTRODUCTION}

It is today widely accepted, that a super massive black hole (SMBH) resides in the very centre of most bright galaxies (Magorrian et al.~1998). 
 Phenomenology of Active Galactic Nuclei (hereafter AGN) can be explained by accretion of material onto the SMBH, with the subsequent emission of radiation, by the unified model (Antonucci 1993). This model explained successfully the occurrence of broad emission lines in polarised light in the spectra of some Seyfert 2 galaxies (Antonucci \& Miller 1985; Miller \& Goodrich 1990), as well as the existence of high-excitation gas extending out from the nucleus with conical and biconical morphology (e.g.~Wilson et al.~1988; Tadhunter \& Tsvetanov 1989; P\'erez et al.~1989). 
One point that remained controversial for a decade is the nature of the blue and ultraviolet continuum found in many Seyfert 2 (Sy2) spectra. This was first attributed to scattered light from the hidden Seyfert 1 (Sy1) nucleus. However, Tran (1995) showed that this cannot be the general case, as the polarisation of the continuum was frequently much lower than that of the broad lines, which cannot be explained by the scattering scenario. Terlevich (1990) proposed a stellar origin for this continuum, and Cid Fernandes \& Terlevich (1995) proposed that a reddened starburst provides this emission. 

The Hubble Space Telescope (HST) has provided a deep insight into the nature of UV light in Seyferts, due to the very high resolution provided, and a relatively high UV sensitivity. High resolution imaging of Sy2 galaxies pointed to a stellar origin for this light in several cases (Heckman et al.~1997; Colina et al.~1997; Gonz\'alez Delgado et al.~1998). Powerful circumnuclear starbursts have been unambiguously identified in 40 per cent of nearby Sy 2 galaxies (Gonz\'alez Delgado et al.~2001; Cid Fernandes et al.~2001, 2004). These starbursts were originally detected by means of either UV or optical spectroscopy of the central few 100 pc. Several spectroscopic works have detected features of young and intermediate age stellar population (Heckman et al.~1997; Gonz\'alez Delgado et al.~1998, 2001; Storchi-Bergmann et al.~2000), suggesting that these populations are significant, if not dominant, in the nuclear region of many Sy2 galaxies.

In order to understand better the nature of the near-UV light in Sy galaxies, as well as the connection between the AGN and starbursts processes, we carried out an HST snapshot survey of the nearest Sy nuclei. The galaxies were imaged with the ACS camera in its High Resolution Channel, through the filter F330W (with an average wavelength of 336.72\,nm). In this way we constructed a useful reference atlas of 75 Sy galaxies which possess HST archival images of their nuclear region, in the near-UV, optical and near-IR, at a very high spatial resolution ($\sim$0.05 arcsec). The sample and the images of the objects, as well as a morphological and photometric analysis, were presented in Mu\~noz Mar\'{i}n et al.~(2007; hereafter MM07). In that work we confirmed the presence of star formation in the circumnuclear region of a fraction of the galaxies ($\sim$60 per cent), and we found that circumnuclear star clusters can be found equally in both Seyfert types. However we have also found extended emission, showing often a biconical structure, and/or bright filaments, which are unlikely to trace star-formation. To investigate the origin of this UV extended emission is the goal of this paper.

There are several processes that are expected to contribute to the extended light in the F330W filter. The most relevant ones are:
\begin{itemize}
\item {\it Stellar continuum}. If compact knots are detected, they are probably young/intermediate age massive star clusters, or clumps of them. But the extended UV emission could also be due to an unresolved population of smaller star-clusters, or generated by the disruption of ageing star clusters (Tremonti et al.~2001; Chandar et al.~2005), as well as emission from the underlying older stellar population of the galaxy.
\item {\it Nebular continuum}. Radiation from free-free and free-bound electron transitions, as well as two photon decay of level 2\,$^2S$ of H\,I, can be an important contribution in regions with ionised gas, and it is expected to be extended, as much of the ionised gas is.
\item {\it Emission lines from ionised gas.} In the wavelength range covered by the F330W filter at the redshift range of the galaxies only two emission lines contribute significantly, [NeV]$\lambda\lambda$3346,3426.
\item {\it Light from the AGN} can be scattered by free electrons or dust and contribute with extended emission.
\end{itemize}

In this work we aim to disentangle the contribution of the different processes that may contribute to the extended near-UV emission. With this purpose, we focus our analysis on a subsample of Seyfert galaxies in which the dominant morphology at near-UV wavelengths is extended emission that is not obviously associated to compact starbursts or star clusters, and for which both near-UV ACS data and WFPC2 [OIII] images exist.

The sample and data are presented in Section \ref{sec:samp}. In Sections \ref{sec:NeV}, \ref{sec:NC} and \ref{sec:scat}, we give an estimation of the expected contribution of each possible component to the extended UV light emission, whereas in Section \ref{sec:UV-oiii} we create synthetic [NeV] and nebular continuum images from a direct scaling of the [OIII] images. In Section \ref{sec:Phot} we explain the photometric analysis performed to study the remaining light after the subtraction of the modeled ionised gas component. A detailed discussion for each object is presented in \mbox{Section \ref{sec:results}}. A general discussion and summary is presented in \mbox{Section \ref{sec:summary}}; and our conclusions can be found in \mbox{Section \ref{sec:conclusions}}.


\section{SAMPLE AND DATA}
\label{sec:samp}

Our sample comprises the galaxies of our Seyfert atlas (MM07) with extended emission, that are also studied by Schmitt et al.~(2003). The latter work presents HST [OIII] images of the nuclear region of 60 nearby Seyfert galaxies, the largest [OIII] imaging compilation of the Narrow Line Region (NLR) of nearby Sy's, which are imaged at a resolution comparable to ours. 
We end up with a final sample of 15 galaxies, from which 8 are Sy2 and 7 are Sy1. For the sake of simplicity we have included the 4 Sy1.5 and the one Sy1.8 (UGC\,12138) in the Sy1 group, as they all have in common a very bright nuclear source that dominates the inner emission at near-UV wavelengths (MM07). The objects, together with some basic properties, are listed in Table \ref{tab:oiii_sample}. There is no significant difference between Sy1 and Sy2 subsamples, neither with respect to their mean distance (71~Mpc for Sy1, and 74~Mpc for Sy2), nor with their mean axial ratio (b/a of 0.80 for Sy1, while it is 0.74 for Sy2). This does not change when considering median values.

The F330W ACS images were reduced by the HST pipeline following the standard procedure of bias subtraction, flat-calibration and distortion correction. Then, the exposures were cleaned from cosmic rays by hand. The [OIII] images were obtained with the WFPC2, with the chip WFC in most cases, and with the PC for some objects; NGC\,3393, was imaged with the WFPC before COSTAR, and had to be deconvolved. The majority of [OIII] images were taken with the linear ramp filter, whose flat-fielding is not performed by the HST pipeline, and requires a bit of extra work. For details on the reduction process of these data, see Schmitt et al.~(2003) for the WFPC2 images, and Schmitt \& Kinney (1996) for NGC\,3393 data.

In addition to the near-UV band, we downloaded and reduced WFPC2 images of the galaxies trough the filter F547M. The use of these images has the advantage of minimising the contamination by strong emission lines, and that all the objects, except for NGC\,7212 and UGC\,1214, are imaged in this band (as it served in most cases for the continuum subtraction of the [OIII] image), what allows for a comparison between them. The images were used for the photometric analysis of Section \ref{sec:Phot}. These images were reduced with the standard procedure using the HST pipeline. When more than one image were available, single exposures were combined and cleaned from cosmic rays with IRAF\footnote{IRAF is distributed by the National Optical Astronomy Observatories, which are operated by the Association of Universities for Research in Astronomy, Inc., under cooperative agreement with the National Science Foundation.} task `crrej'. Remaining cosmic rays were eliminated by hand. Finally, the images were resampled to the ACS scale.

We completed the photometric set with NICMOS F160W images. This last filter has the advantage that there are data for all but one (NGC\,7212) of the objects, and that it is so wide that we can neglect the contribution from emission lines. NICMOS single exposures were corrected for the `pedestal effect' (produced by changes in the bias levels of the different chips during the exposure) with IRAF task `pedsub', combined with the standard calibration pipeline, and resampled to ACS resolution.

Fig.\,1 shows a comparison between the near-UV ACS images and the narrow-band continuum subtracted [OIII]$\lambda$5007 images.
 This comparison shows a very similar morphology in most cases. In some cases, especially in those of clear bi-conical structure, the morphology is identical to that of the [OIII] (IC\,5063, NGC\,3393, and UGC\,1214), implying that the near-UV emission is produced in the same region that the ionised gas, and thus nebular continuum and [NeV] emission will be the main contributors to the UV light emission. 
 In the case of NGC\,3516 the morphology is very dissimilar, with an extended UV emission surrounding the bright nucleus that does not match the bi-conical structure in [OIII]. This means that in this case we may be observing mostly light which seems not to be associated to the ionised gas emission. Also in NGC\,5347 this kind of emission is likely responsible for the conical region extending to the north at less than 1~arcsec from the central source. There are also a few cases in which, in addition to the extended emission, the UV light comes from knots or clumps produced probably by star clusters. NGC\,4253 and NGC\,7212 are two examples of this.

 \begin{figure*}
\label{fig:oiii}
  \centering
  \begin{minipage}{140mm}
   \includegraphics[angle=0,width=\textwidth]{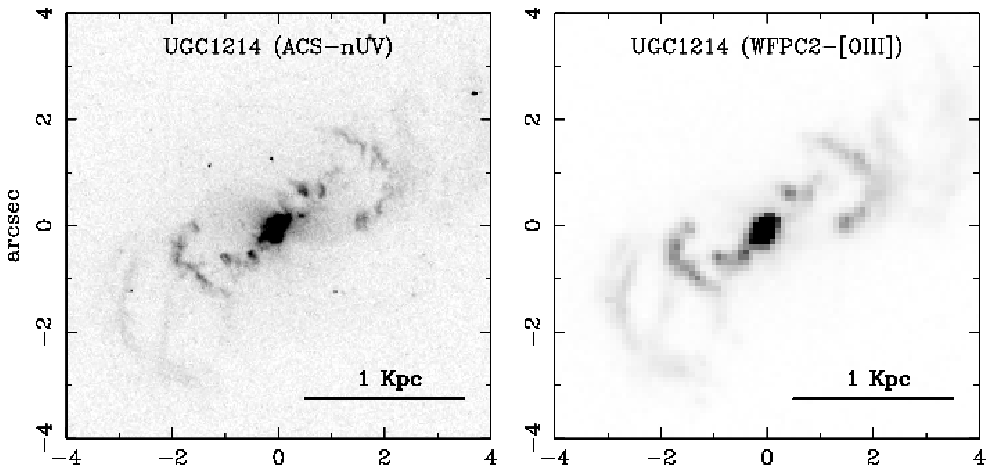}
   \includegraphics[angle=0,width=\textwidth]{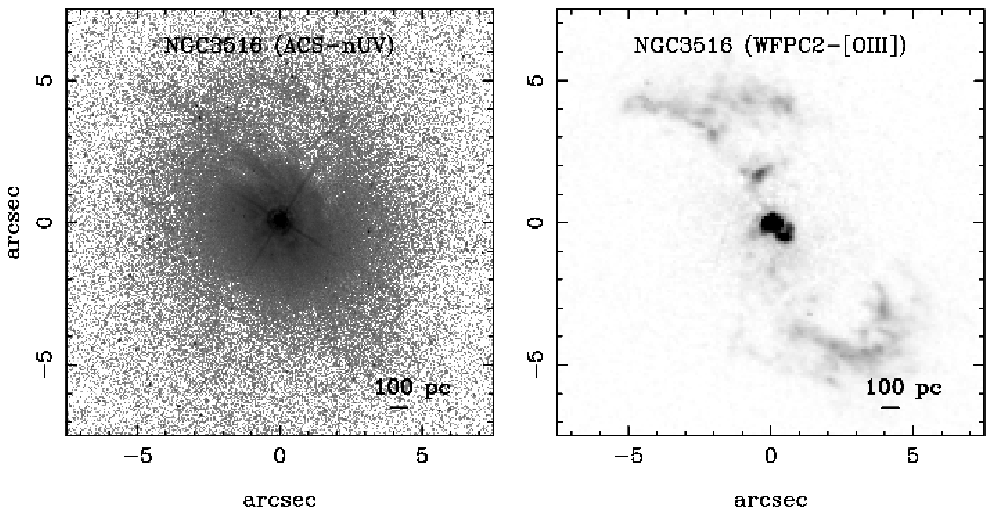}
   \includegraphics[angle=0,width=\textwidth]{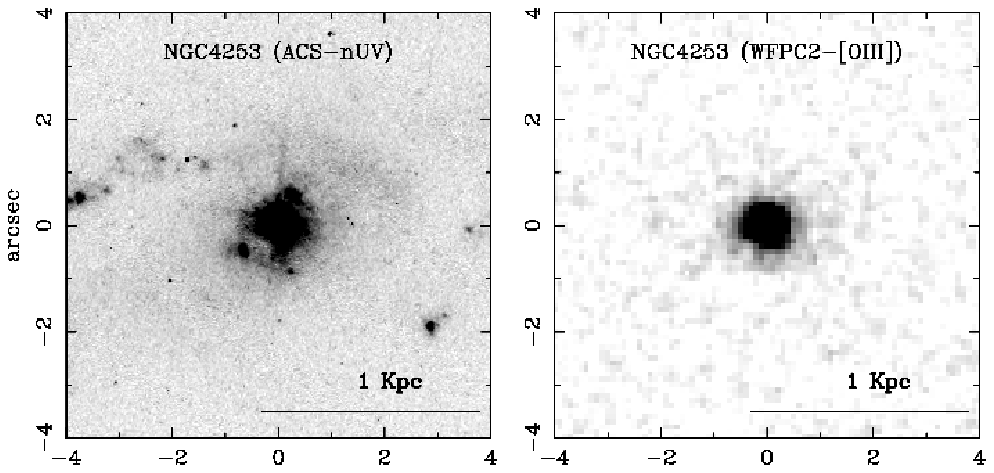}
   \caption{Three examples comparing the UV and ionised gas morphology. On the left, the images in F330W with ACS; on the right, images in [OIII]$\lambda$5007 with WFPC2. The region of interest is plotted with the same scale for each object. For UGC\,1214 the morphology at near-UV is identical to that of the ionised gas, while for NGC\,3516 it is very dissimilar. NGC\,4253 constitutes an example in which UV extended emission appears with compact sources (probably SC), but the knots are not detected in the [OIII] image}
   \end{minipage}
\end{figure*}

\begin{table*}
  \centering
  \begin{minipage}{165mm}
   \caption{Basic data and line ratios for the galaxies of the sample. }
   \begin{tabular}{@{}lcccccccccl@{}}
   \hline
    Object   & Alternative & Activity & Distance & Scale  & B$_T$  & E(B-V)  &  b/a  &  [OIII]/[NeV]  &  [OIII]/H$\beta$  &  Ref.\\
            &   name       & type    &  (Mpc)  & (pc/'') & (mag)  &        &       &     &        &   \\
   \hline
IC\,5063     & ESO\,187-G23  &  Sy2     & 44.8  & 217  & 12.89  & 0.061  & 0.67  &       &   8.8    & dG92\\
Mrk\,6       & IC\,450       & Sy1.5    & 80.8  & 392  & 15.0   & 0.136  & 0.625 &  12.0 &  2.4$^a$ & mo83\\
Mrk\,915     &               &  Sy1     & 99.0  & 480  & 14.82  & 0.063  & 0.833 &       &  7.0$^a$ & dG92 \\
NGC\,1320    & Mrk\,607      &  Sy2     & 35.3  & 171  & 13.32  & 0.047  & 0.316 &       &   9.0    & dG92 \\
NGC\,3393    &               &  Sy2     & 50.3  & 244  & 13.09  & 0.075  & 0.909 &  16   &  12.5     &  sb96 \\
NGC\,3516    &               & Sy1.5    & 40.8  & 198  & 12.5   & 0.042  & 0.765 &  7.5  & 1.6$^a$  &  an70 \\
NGC\,4253    & Mrk\,766      & Sy1.5    & 56.2  & 273  & 13.70  & 0.020  & 0.8   &  5.7  &  13      & gp96b \\
NGC\,4593    & Mrk\,1330     &  Sy1     & 38.8  & 188  & 11.67  & 0.025  & 0.744 &       & 0.33$^a$ & dG92 \\
NGC\,5347    &               &  Sy2     & 36.6  & 178  & 13.4   & 0.021  & 0.765 &  6.2  &  10.0  &  STIS \\
NGC\,5548    &               & Sy1.5    & 74.0  & 359  & 13.3   & 0.020  & 0.929 &  3.3  &   8.1    &  cr98  \\
NGC\,7212    &               &  Sy2     & 111.1 & 538  & 14.78  & 0.072  &       &       &   10.8   &  wh92 \\
NGC\,7674    & Mrk\,533      &Sy2-H\,II & 119.7 & 580  & 13.92  & 0.059  & 0.909 &  6.8  &  14.3    &  STIS \\
UGC\,1214    & Mrk\,573      &  Sy2     & 70.0  & 339  & 13.68  & 0.023  & 1.0   &  5.6  &  14.6    &  STIS \\
UGC\,6100    &               &  Sy2     & 124.0 & 601  & 14.30  & 0.012  & 0.617 &       &   12.0   &  cg94 \\
UGC\,12138   &               & Sy1.8    & 103.9 & 503  & 14.24  & 0.085  & 0.875 &       & 2.71$^a$ &  dG92\\
 \hline
 \label{tab:oiii_sample}
 \end{tabular}
 
Col.(1,2)\,: galaxy name; Col.\,(3): spectral type; Col.\,(4): distance corrected from a Virgo Infall model; Col.\,(5): physical scale of the images; Col.\,(6): total asymptotic magnitude in B, B\_T, from RC3 catalogue; Col.\,(7): reddening, E(B-V); Col.\ (8): axial ratio (b/a); Col.\,(9): ratio [OIII]$\lambda$5007/[NeV]$\lambda\lambda$3346,3426; Col.\,(10): ratio [OIII]$\lambda$5007/H$\beta$; Col.\,(11): references for emission line fluxes (an70: Anderson 1970; cr98: Crenshaw et al.~1998; cg94: Cruz-Gonz\'alez et al.~1994; dG92: de Grijp et al.~1992; gp96a: Gonz\'alez Delgado \& P\'erez 1996a; gp96b: Gonz\'alez Delgado \& P\'erez 1996b; mo83: Malkan \& Oke 1983; wh92: Whittle 1992; STIS: calculated from STIS nuclear spectra) 
\\ Quantities for columns 2 to 8 are extracted from NED.
\medskip

 $^a$ Includes H$\beta$ broad component.
 \end{minipage}
 \end{table*}


\section{ANALYSIS OF THE NEAR-UV AND [OIII] IMAGES}


\subsection{Estimation of the [Ne\,V] lines contribution}
\label{sec:NeV}

The forbidden lines [NeV]$\lambda\lambda$3346,3426 are the only emission lines strong enough to contribute to the light through the F330W filter. These high ionisation lines are expected to be strongly diluted in the presence of a bright continuum and give a rather small equivalent width. Studying a sample of NLSy1 of the ROSAT Deep Survey, Hasinger et al.~(2000) find that the equivalent width of [NeV] is only 4\,\AA, probably as a consequence of the strong continuum contribution. However [NeV] lines can also be locally strong, in the presence of a hard ionising spectrum. For example, Soifer et al.~(1995) find strong [NeV] lines in FSC\,10214+4724, a Sy2 and ULIRG at high redshift, with a [OIII]/[NeV] ratio of 3.85. Storchi-Bergmann et al.~(1996) find that this ratio can vary from 2 to 30, although most objects fall in the range [OIII]/[NeV]$\sim$5-15. Note the caveat that most of these measurements come from the study of the NLRs, while the emission in our sample extends in many cases to several arcseconds, corresponding to the Extended Narrow Line Region (ENLR).

In Table \ref{tab:oiii_sample}, we list [OIII]/[NeV] and [OIII]/H$\beta$ ratios for the objects of our sample. These have been compiled from a variety of sources. We found published fluxes for [NeV] for a limited number of galaxies (Mrk\,6, NGC\,3393, NGC\,3516, NGC\,4253, NGC\,5347, NGC\,5548). In order to complement these data, for the objects NGC\,3393, NGC\,5347, NGC\,7674, and UGC\,1214 (Mrk\,573), we have used published spectra from Spinelli et al.~(2006), obtained with HST/STIS and with a spectral range covering the [NeV] wavelength. These spectra have been extracted in windows of 0.2 arcsec and cover the optical and UV. This window is the best possible angular resolution achievable with the observations, corresponding to a region of 52 pc radius for NGC3393, 120 pc for NGC7674, 33pc for NGC5347 and 72pc for Mrk573 centred on the nucleus. For details of the reduction process of the spectra see Spinelli et al.~(2006).

 The line fluxes in the STIS spectra were measured with IRAF task `splot' through a gaussian fitting to the line.
 For the two objects with STIS spectra and published [NeV] data (NGC\,3393 and NGC\,5347), we find a reasonable agreement between our measurements and published data. The ratio [OIII]/H$\beta$ is around 10 in both cases, while in the ratio [OIII]/[NeV] there is a factor 2 difference, which is nevertheless within the expected uncertainty in the flux measurements of [NeV] lines. In the case of NGC\,5347 we will use the STIS value, as the data in Gonz\'alez Delgado \& P\'erez (1996a) is very noisy in the UV, while for NGC\,3393 we will use the data of Storchi-Bergmann et al.~(1996), as the STIS spectra is noisier in this case. Overall, [OIII]/[NeV] ratios vary between 3 and 20.

Due to the lack of data of [NeV] for some galaxies of the sample, we use bibliographic data of other objects in order to constrain their expected range of line ratio values. In Table \ref{tab:NeVliterature} we compile some values of [OIII]/[NeV] and [OIII]/H$\beta$ for galaxies with different activity types, found in several studies (Storchi-Bergmann et al.~1995; Morris \& Ward 1988; and Koski 1978). Overall, the objects and the compiled data, are very inhomogeneous. 
However, this should give us an estimate of the expected value of [OIII]/[NeV] ratio. This line ratio can be used to scale the narrow-band [OIII] image, in order to obtain an approximate [NeV] image. 
 For Sy1 (Sy1, 1.2 and 1.5) we find a mean [OIII]/[NeV] ratio of 8.4, with a dispersion of 5.8; for Sy1.9 galaxies this value is 22.2 with 9.1 dispersion; 
 and for Sy2 galaxies we calculate a ratio of 16.9 with a dispersion of 6.7. We want to remark the wide range for the ratios, from 1 to a few tens. Also, the values from Storchi-Bergmann et al.~(1995) may be upper limits to the actual NLR value, as they used very wide slits in order to match the aperture of IUE data which was used in combination with the optical data. From these results it can be concluded that any scaling between 1/2--1/30 to the [OIII] image may be a reasonable approximation to the [NeV] emission, depending very much on the nature of the object. Note, however, that these values have been obtained from spectra dominated by the NLR contribution, and we are assuming that the [OIII]/[NeV] ratio in the ENLR is similar to the value in the NLR.

 \begin{table*}
  \centering
  \begin{minipage}{110mm}
   \caption{[NeV] data for galaxies in the literature.}
   \begin{tabular}{@{}lcccccccl@{}}
   \hline
    Name     &    Redshift  & Type &   [OIII]/[NeV]  &  [OIII]/H$\beta$  &  Ref.\\
   \hline
1ES\,1615          & 0.0379  & Sy1   & 20.3 &    1.4   &  mw88 \\
3C\,33             & 0.0597 & NLRG/Sy2 & 24.6 &  13.3  &  ko78 \\
3C\,184.1          & 0.1182  & RG/Sy1 & 16.3 &  11.1     & ko78 \\
3C\,273            & 0.1592  &  QSO  & 1.9  &   0.13     &  mw88 \\
3C\,445            & 0.0568  &  Sy1  & 13.0 &    2.0   &  mw88 \\
IC\,3639           & 0.0109  &  Sy2  & 11.6 &    5.3   &  sb95 \\
IRAS\,F18389-7834  & 0.0742  &  Sy1  & 2.1   &   0.28    &  mw88 \\
Fairall\,51        & 0.0138  &  Sy1  & 4.6  &   0.89   &  mw88 \\
MCG\,-6-30-15      & 0.008   & Sy1.2 & 10.9 &   0.64   &  mw88 \\
MKN\,509           & 0.0329  & Sy1.2 & 0.9  &   0.35   &  mw88 \\
MKN\,841           & 0.0365  & Sy1.5 & 5.0  &   0.57   &  mw88 \\
MKN\,871           & 0.0333  & Sy1.5 & 4.6  &   0.73   &  mw88 \\
MKN\,896           & 0.0262  &  Sy1  & 4.5  &   0.32   &  mw88 \\
MRK\,1393          & 0.0543  & Sy1.5 & 13.1 &   1.2    &  mw88 \\
NGC\,4748          & 0.0146  &  Sy1  & 18.4 &   6.0    &  sb95 \\
NGC\,3783          & 0.0096  &  Sy1  & 10.7 &   0.86   &  mw88 \\
NGC\,4593          & 0.0087  &  Sy1  & 6.9  &   0.30   &  mw88 \\
NGC\,449           & 0.0159  &  Sy2  & 24.0  &  11.8   &  ko78 \\
NGC\,1068          & 0.0038  &  Sy2  & 6.7   &  13.0   &  sb95 \\
NGC\,3393          & 0.0125  &  Sy2  & 27.6 &   11.0   &  sb95 \\
NGC\,3081          & 0.0080  &  Sy2  & 14.4 &   14.3   &  sb95 \\
NGC\,5506          & 0.0059  & Sy1.9 & 35.1 &   7.2    &  mw88 \\
NGC\,5643          & 0.0038  &  Sy2  & 12.5   & 13.1   &  mw88 \\
NGC\,5940          & 0.0339  &  Sy1  & 1.8  &   0.36   &  mw88 \\
NGC\,6814          & 0.0051  & Sy1.5 & 8.6  &   0.77   &  mw88 \\
NGC\,7314          & 0.0054  & Sy1.9 & 16.5 &   8.6    &  mw88 \\
TOL\,113           & 0.0571  & Sy1.9 & 14.9 &   9.5    &  mw88 \\
TOL\,20            & 0.0233  & Sy1   & 7.4  &   0.86   &  mw88 \\
UGC\,10683b        & 0.0308  & Sy1   & 1.6  &   0.59   &  mw88 \\
\hline
 \label{tab:NeVliterature}
 \end{tabular}

Col\,(1): Object name; Col\,(2) Redshift; Col\,(3) Activity type; Col\,(4) Measured line ratio for  [OIII]$\lambda\lambda$5006.8 over [NeV]$\lambda\lambda$3425.8; Col\,(5) Line ratio for  [OIII]$\lambda\lambda$5006.8 over H$\beta$; Col\,(6) References: mw88, Morris \& Ward (1988); sb95, Storchi-Bergmann et al.~(1995); and ko78, Koski (1978). 

\end{minipage}
\end{table*}



\subsection{Estimation of the nebular continuum emission}
\label{sec:NC}
\index{sec:NC}
Due to the short wavelength and large width of the near-UV filter, the nebular continuum may be an important contribution at this band. The main contributions to this continuum are electron-ion recombinations. There is also a contribution of free-free transitions, by means of free electrons-ion interaction followed by bremsstrahlung radiation from the electrons. The latter is more important at longer wavelengths, and is negligible in the F330W filter band. The most important role is played by hydrogen and helium recombination.  
Another source of nebular continuum at this wavelength is the two-photon decay of the 2$^2S$ level of H\,I. 

The emission coefficient for the mechanisms described above depend both on frequency and temperature, and can be written as:
\begin{align}
&j_\nu(H\,I)    =\frac{1}{4\pi}N_pN_e\gamma_\nu(H^0,T)\\
&j_\nu(He\,I)   =\frac{1}{4\pi}N_{He^+}N_e\gamma_\nu(He^0,T)\\
&j_\nu(He\,II)  = \frac{1}{4\pi}N_{He^{++}}N_e\gamma_\nu(He^+,T)\\
&j_\nu(2q)    = \frac{1}{4\pi}N_{p}N_e\gamma_\nu(2q)
\end{align} 
where $\gamma_\nu$ is a coefficient that takes into account the bound-free and free-free transitions for each ion at a given temperature. The coefficient $\gamma_\nu$ is calculated numerically. We used the tabulated data of Ercolano \& Story~(2006).

In order to calculate the total intensity of radiation ($I$), $j_\nu$ has to be integrated over the solid angle, frequency range, and optical depth. In order to get rid of the geometry, it is more useful to relate the continuum intensity to the intensity of some important emission line, often Balmer H$\beta$, whose emission coefficient ($j_{H\beta}$) is defined as:
\begin{equation}
N_pN_e\alpha_{H\beta}^{eff} = \frac{4\pi j_{H\beta}}{h\nu_{H\beta}}
\end{equation}
where $\alpha_{H\beta}^{eff}$ is the effective recombination coefficient.

So the relation between intensities is:

\begin{equation}
\frac{I_\nu(ion)}{I_{H\beta}} = \frac{N_{ion^+}}{N_p}\frac{\gamma_\nu(ion,T)}{h\nu_{H\beta}\alpha_{H\beta}^{eff}}
\end{equation}

And taking into account the conversion between frequency and wavelength, the wavelength intensity can be written in terms of I$_{H\beta}$ as:

\begin{equation}
I_\lambda(ion) = \frac{N_{ion^+}}{N_p} \frac{\lambda_{H\beta}}{\lambda^2 h\alpha_{H\beta}^{eff}} \gamma_\nu(ion,T) I_{H\beta}
\label{eq:Flam(ion)}
\end{equation}

From a direct comparison of the $\gamma$ factors, in Fig 4.1 of Osterbrock (1989), we obtain that the contribution of two-photon continuum will be $\sim$30 per cent of that of H\,I recombination (half a dex smaller). 
In order to asses the contribution to the nebular continuum from the recombination of each different ion, we need to know the ionisation fraction of each species. We obtained the mean ionisation running the CLOUDY code (Ferland et al.~1998) with typical NLR values, such as a Lexington model (Ferland 1995), which has electron density of 10$^4$~cm$^{-3}$ and temperature 10$^4$~K. The models yield a similar ionisation fraction for H and He. From the ionised He, 70 per cent is singly ionised and 30 per cent twice ionised.

For He\,I, the factor $\gamma_\nu(He\,I,T)$ has a break at 342.2\,nm, longward of which it falls approximately an order of magnitude. This is almost in the middle of the F330W filter. In order to calculate how it affects the nebular continuum contribution we integrate the filter throughput considering that, for wavelengths larger than 342.2\,nm, the contribution is 10 per cent the one for smaller wavelengths and then divide by the total integrated troughput. 
This ratio is 0.66 at restframe, but rises to 0.75 for the objects with the highest redshift in the sample. We will use a mean correction of 0.7 for the continuum estimation.
Therefore we can use the next formulae which relate the continuum fluxes with the helium abundance (Y) and the ionisation fraction (IF):
\begin{equation}
\frac{I_{NC}(He\,I)}{I_{NC}(H\,I)} = {IF(He\,I)}{Y} \frac{\gamma_\nu(He\,I,T)}{\gamma_\nu(H\,I,T)} 0.7
\end{equation}
\begin{equation}
\frac{I_{NC}(He\,II)}{I_{NC}(H\,I)} = {IF(He\,II)}{Y} \frac{\gamma_\nu(He\,II,T)}{\gamma_\nu(H\,I,T)}
\end{equation}

The relation between $\gamma_\nu(He\,I,T)$ and $\gamma_\nu(H\,I,T)$ at $\sim$330\,nm is approximately equal to 1.2. The ratio for He\,II recombination is 2.4. Assuming an abundance ratio of 1/10, and an IF of 0.7 for He$^+$ and 0.3 for He$^{++}$, the contributions of He\,I and He\,II recombination to the nebular continuum are $\sim$6 per cent and $\sim$7 per cent respectively of that of H\,I.
We will assume a temperature for the ionised gas of 10,000~K. In order to calculate the contribution in flux of the nebular continuum to the F330W filter, we calculate $I_\lambda(H\,I,T)$ at the average wavelength of the filter (AVGWV) and T=10,000\,K and multiply by its rectangular width (RECTW). The parameters AVGWV and RECTW of the filter F330W were calculated with IRAF SYNPHOT task `bandpar', yielding 336.72\,nm ~and 54.418\,nm ~respectively.
As a good approximation, the nebular continuum flux in the filter F330W can thus be estimated as follows:
\begin{equation}
I_{NC,F330W} \simeq 1.43 \cdot I_{AVGWV}(H^0,T)\cdot RECTW ,
\end{equation}
where the factor 1.43 corresponds to the sum of the contribution of the H\,I recombination continuum (1), the two-photon continuum (0.30), and the He\,I and He\,II recombination continuum (0.06 + 0.07).

And from Equation~\ref{eq:Flam(ion)} and the values of the parameters AVGWV, RECTW and T, we obtain:
\begin{equation}
I_{NC,F330W} \simeq 3 \cdot I_{H\beta}
\label{eq:FNC}
\end{equation}

In our case it is more convenient to express the former relation in terms of the [OIII] flux. This will introduce a further uncertainty, but it is a necessary step, in order to adapt the study to the data we have. We expect that the narrow line components of both Sy1 and Sy2 have similar [OIII]/H$\beta$ line ratios, when only the narrow component of H$\beta$ is included. 
For the Sy2 galaxies in Table \ref{tab:NeVliterature}, we calculate a mean [OIII]/H$\beta$ ratio of 12.3 with a dispersion of 2.8, this being 8.61 and 0.87 respectively for Sy1.9. For the Sy1 galaxies, Morris \& Ward (1988) tabulate the total H$\beta$ line flux, including the broad component. Including the broad component would yield an unrealistically high nebular continuum for the NLR. The data for the object 3C\,184.1, from Koski (1978), include only the narrow component, and the ratio is 11.1. This value is consistent with those from Sy2, and also agrees with the data in Table \ref{tab:oiii_sample}, in which [OIII]/H$\beta$ ranges from 8 to 14. Therefore we will assume a single [OIII]/H$\beta \simeq 10$ for all the Sy galaxies.

 With these considerations, and the caveats expressed above, we can make a prediction of the nebular continuum through F330W in terms of the [OIII] flux with the equation:
\begin{equation}
I_{NC,F330W} \sim f \cdot I_{[OIII]},
\label{eq:FNC2}
\end{equation}
with {\bf f}~=~0.3 and taking into account that, for the range of the line ratio cited above, {\bf f} may vary between 0.375 and 0.2.

Dust obscuration may have an important effect in the calculated ratio between [OIII] and near-UV emission. Extinction coefficients for the ACS filters using the interstellar extinction from Cardelli et al.~(1989) are tabulated in Sirianni et al.~(2005), from where we take $A_{F330W}/E(B-V)=5.054$, and A$_{F502N}/E(B-V)=3.458$. Thus, the difference in magnitude between filters F330W and F502N due to obscuration is 
\begin{equation*}
A_{F330W}-A_{F502N}=1.6 E(B-V)=1.6 A_V/3.1 \simeq 0.5 A_V ,
\end{equation*}
what translates into a flux ratio of $10^{-0.2 A_V}$. This means that an extinction of $A_V=1$, would reduce the near-UV contribution of the ionised gas emission to 63 per cent of its unextincted value, reducing the value of {\bf f} in Equation~\ref{eq:FNC2} from 0.3 to 0.19. Thus,
\begin{equation}
F_{NC,F330W} \sim 0.3 \cdot 10^{-0.2 A_V} \cdot F_{[OIII]},
\label{eq:FNC3}
\end{equation}
where we have adopted 0.3 as an average scaling for the [OIII] images, keeping in mind that the reddening effect may change its value.

The value of {\bf f} is subject to several other effects and assumptions. From the values of $\gamma_\nu(ion,T)$ tabulated in Osterbrock (1989), we estimate that a change of 1000\,K in the temperature may produce a change up to 10 per cent in the value of {\bf f}. The ionisation parameter (U) can also affect the result. In the regions of highest U there might be a higher contribution of the nebular continuum. The ratio [OIII]/H$\beta$ is expected to vary with U, although data from Table \ref{tab:oiii_sample} suggest a quite stable ratio for these objects. Anyway, a certain degree of spatial variation of the scaling factor {\bf f} is to be expected.
In addition, as noted by Luridiana et al.~(2003), the NC from two-photon decay could be higher due to the reprocessing of Ly$\alpha$ photons inside the nebula. For a theoretical thick and dust free nebula, the contribution of two-photon decay could be 3 times higher than the one considered, becoming as important as hydrogen recombination, and rising the value of {\bf f} in 
equation~\ref{eq:FNC} from 3 to 5. However the latter value is an upper limit, and the actual value should lie in between.
Due to all these caveats, this number has to be taken just as a reference value.



\subsection{Contribution of light scattered from the AGN}
\label{sec:scat}

On basis of the unified model we expect this contribution in both Sy types. This emission is expected to correlate spatially with the [OIII] emission to a certain degree, although the correspondence may not be perfect. 
Scattered light can be identified through polarimetric studies, as it is expected to be at least partially polarised. The two processes have a different dependence on density N$_e$, and polarisation also depends on the scattering angle; line emission would roughly go with N$_e^2$ while polarised flux would go with N$_e\times$P, where the degree of polarisation (P) depends on the scattering angle. To make things more complicated, a low polarisation does not necessarily imply a low contribution from the scattered light, as the scattered light might have a low intrinsic polarisation.
 
Scattered nuclear emission from dust is known since long in some bona-fide hidden Sy1, as NGC\,1068, in which it occurs in compact knots and extended regions far from the nucleus (Antonucci \& Miller 1985; Miller, Goodrich, \& Mathews 1991). In some objects, the UV emission is known to be dominated by scattered light from the AGN. Capetti et al.~(1995), find that most, if not all, of the central UV emission in NGC\,1068 may be due to scattered light from the obscured AGN (polarisation up to 65 per cent). Kishimoto et al.~(2002a) find a polarisation of $\sim$20 per cent for Mrk\,3, which has an extended emission with biconical clumpy morphology similar to some objects in our sample (UGC\,1214, NGC\,3393 or IC\,5063). On the other hand, Kishimoto et al.~(2002b) study Mrk\,477 concluding that scattered light is responsible for only as much as 10 per cent of the UV emission. Smith et al.~(2002) study a sample of Sy1 galaxies, several of which are in our sample, finding a degree of polarisation bellow 2 per cent at optical wavelengths. Nevertheless, the polarisation value could be higher, if one considers that it may be diluted by the contribution of the nuclear flux. We have estimated the contribution of the nuclear flux in our Sy1 images by subtracting a PSF (point-spread function) obtained from the image of a star observed through the same filter. On average, we had to scale the PSF to 75 per cent of the peak nuclear flux in order to eliminate the nuclear point source.
 Then, assuming that the bulk of the polarisation comes from the extended emission, and considering that the AGN emission accounts for 3/4 of the total nuclear emission, the degree of polarisation would be around 8 per cent. That could imply an important contribution of the scattered light, in case that the intrinsic polarisation of this component were around 10 per cent.

Inglis et al.~(1993) studied IC\,5063, measuring a polarisation of 1-2 per cent in the optical continuum. They argue that scattered light cannot account for most of the emission. NGC\,7212 and NGC\,7674 are traditionally considered highly polarised objects (Miller \& Goodrich 1990; Kay 1994), so they are likely to have a strong contribution of scattered light. By fitting a galactic template to some spectral features, Tran (1995) calculated a contribution of the stellar continuum of 73 per cent and 56 per cent, respectively. However, as they discuss, this determination is very uncertain, and it depends of the determination method (see, for example, Kay 1994).

As a summary, we could say that the contribution of scattered light in the circumnuclear region may be important for some objects, although it is not trivial to correct for it. A detailed study of this component would need polarimetric imaging, that is beyond the scope of this work.




\subsection{Analysis: contribution of the ionised gas to the near-UV light}
\label{sec:UV-oiii}
In the previous sections we have shown that the contribution of [NeV]$\lambda\lambda$3346,3426 emission line fluxes and the nebular continuum to the light in the filter F330W, is expected to be of the order of 30 per cent of the [OIII]$\lambda$5007 flux for Sy galaxies. However, this contribution may vary from less than 20 per cent, up to 70 per cent in extreme cases, depending on the factors explained above. 
 Both components originate in the same region, most probably the photoionised NLR and ENLR, therefore their effect is reinforced in the image. Our aim in this section is to check whether the contribution of the [NeV] lines and the nebular continuum are enough to explain the similarities of the morphology of the objects in the near-UV F330W and in the [OIII] narrow band images. We do this by creating a synthetic ionised gas ([NeV] emission + nebular continuum) image, by direct scaling of the [OIII] image.

The [OIII] images were obtained with WFPC2, and have a lower resolution, so first we had to resample these images to the ACS resolution with the IRAF task `geotran'. For every pair of near-UV and [OIII] images, we convolved each image with the PSF of the other instrument configuration, which we created using the software TinyTim. For the [OIII] images taken with the WFPC2 linear ramp filter, a monochromatic wavelength of 500.7~nm was used when creating the PSF.
We calibrated the ACS images into integrated flux, by multiplying by the inverse sensitivity (header keyword PHOTFLAM) and the filter width (RECTW parameter) calculated by the task `bandpar' of the SYNPHOT package. Finally, we aligned and trimmed the images using some sharp features in the image. The result of this process are flux calibrated [OIII] and near-UV images that can be directly compared.

All the Sy1-1.8 galaxies in the sample show a very bright compact nuclear source. This source is unresolved in the ACS frames (MM07), showing the features of a pure PSF in the images, such as diffraction spikes and Airy rings. Before comparing the [OIII] and the ACS images, we performed a subtraction of the compact nuclear source from the ACS frames using a bright star, with a colour similar to that of the galaxies, and observed with the same instrumental configuration, as PSF template\footnote{ACS PSF Characterisation HST Proposal 9667}.
For each galaxy with an obvious diffraction pattern, the star image was scaled to match the intensity of the central source of the galaxy, in such way that each galaxy had its own reference PSF. This was done by shifting the position of the peak flux in the PSF image to that of the compact nucleus of each galaxy and scaling it to the same flux value. Then the reference PSF was scaled by a factor of 5 per cent and a series of images were created by subtracting from the galaxy image this scaled reference PSF in steps of 5 per cent. Finally, for each galaxy, we ended up with several images, each one corresponding to the subtraction of  the scaled PSF by increasing factors of 5 per cent, 10 per cent, 15 per cent... The best subtraction was chosen by visual inspection, choosing the image where the diffraction spikes had completely disappeared and the nuclear residuals near the centre, which inevitably appear, were minimised. These residuals are mostly due to the undersampling of the images, breathing effects (slightly change of focus) of the telescope optics and due to alignment errors of about 0.05 pixels.
The whole process is performed in the \_FLT images (before correcting from geometric distorsion) and then the resulting images were distorsion-corrected. This proved to leave smaller residuals than working directly in the final \_DRZ images (after multidrizzle task correction). 
Further details on this procedure will be given in a forthcoming paper about the morphology of the central regions of Sy1 galaxies (Spinelli et al., in preparation). 
After this procedure we were left with near-UV images free from the nuclear contribution, which can now be compared with the [OIII] images, after the matching procedure described in the previous paragraph.

For each object, we have next scaled the [OIII] image in order to create a synthetic nebular continuum plus [NeV] image. We expect there might be variations of the ratio [OIII]/[NeV] over the field of view, but as a first approximation we neglect this position dependence. 
As described in the previous section, {\bf f} is the scale factor we use to multiply the [OIII] image previous to its subtraction from the flux-calibrated near UV image (F330W image in flux density times RECTW). We have tried different {\bf f} values, looking for the one that best removes the extended emission. During this process we have also refined the alignment of the images by minimising the residuals.
In previous sections, we have discussed the range of reasonable {\bf f} values (0.2--0.7, but most frequently expected $\sim$0.3). Based on that we can deduce for each object individually, whether the ionised gas can explain the morphology observed in the near-UV exposure. 
Using this procedure we have created the synthetic [NeV]+NC images, and estimated the {\bf f} factor for each object. This is illustrated in Fig.~\ref{fig:all3_01}, in which we show the synthetic images with the most likely value of {\bf f}. In some cases we use a very high {\bf f} in order to show that the ionised gas cannot account for most of the emission. The result is commented in detail for each object in the next section, including the photometric analysis of the residuals, which we discuss bellow.

\begin{figure*}
%
  \centering
  \begin{minipage}{160mm}
\includegraphics[angle=-90,width=\textwidth]{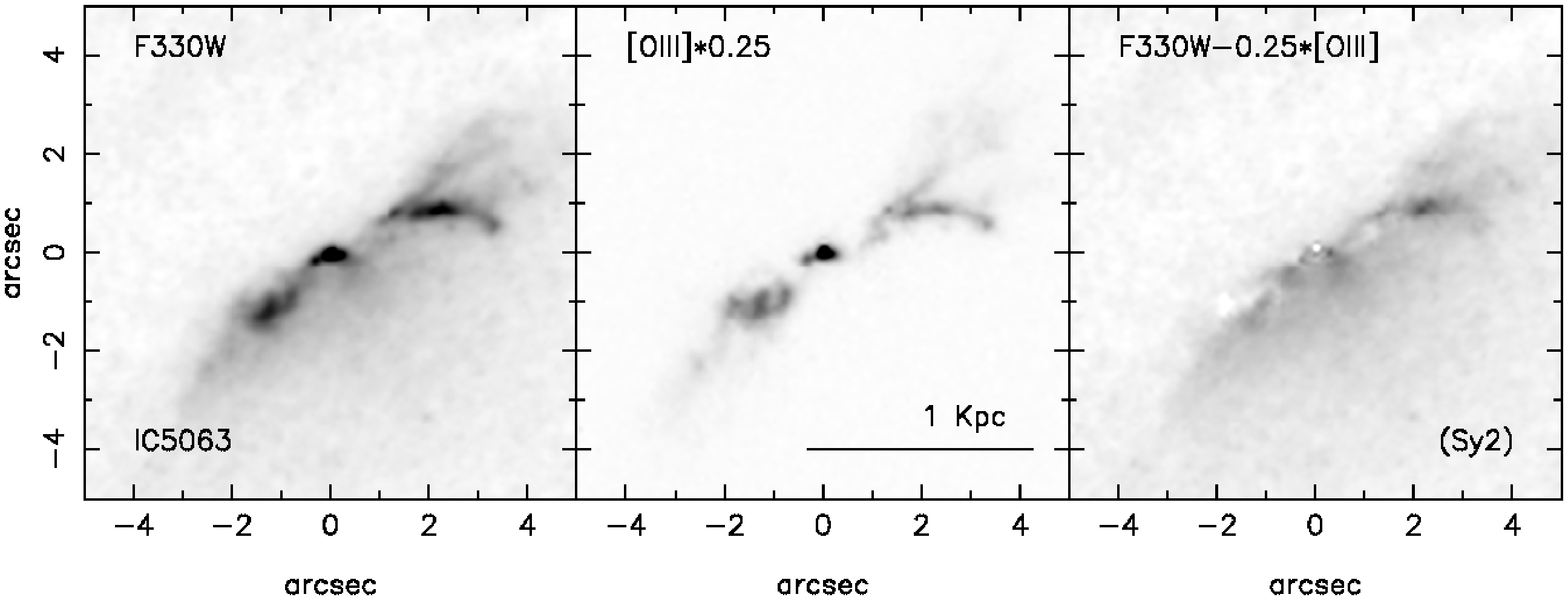}
\includegraphics[angle=-90,width=\textwidth]{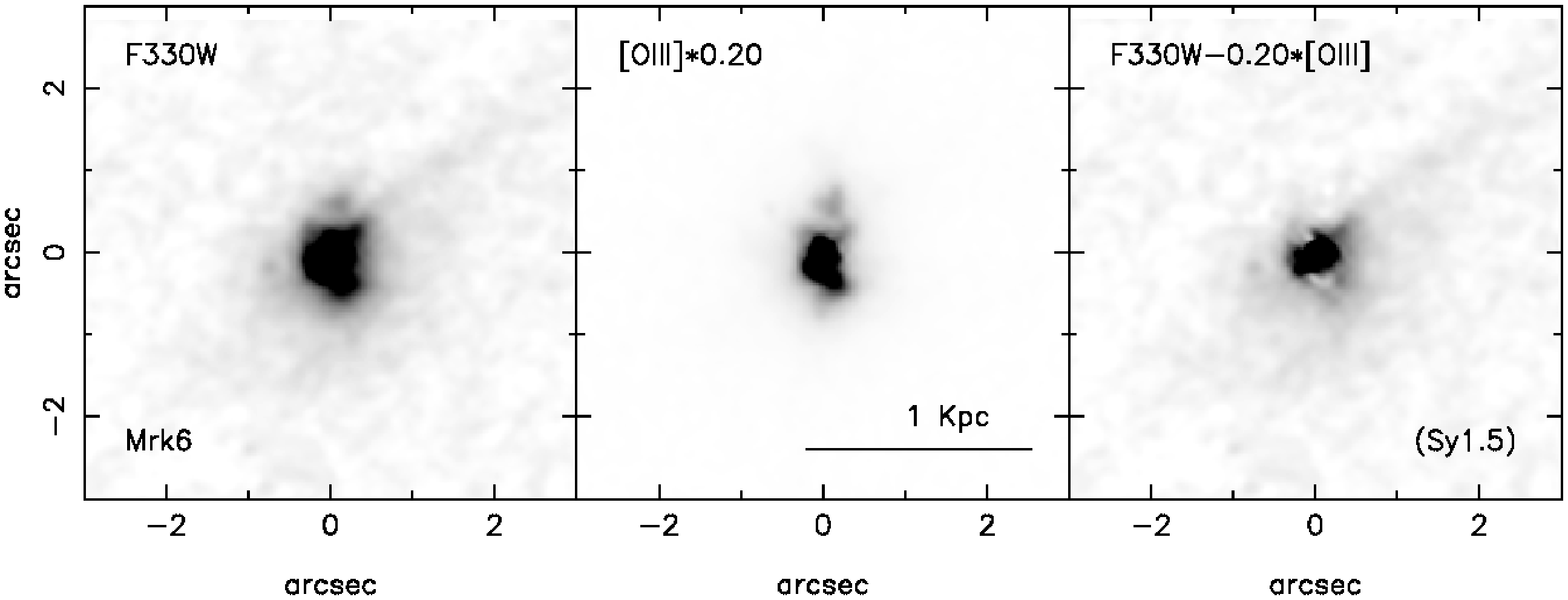}
\includegraphics[angle=-90,width=\textwidth]{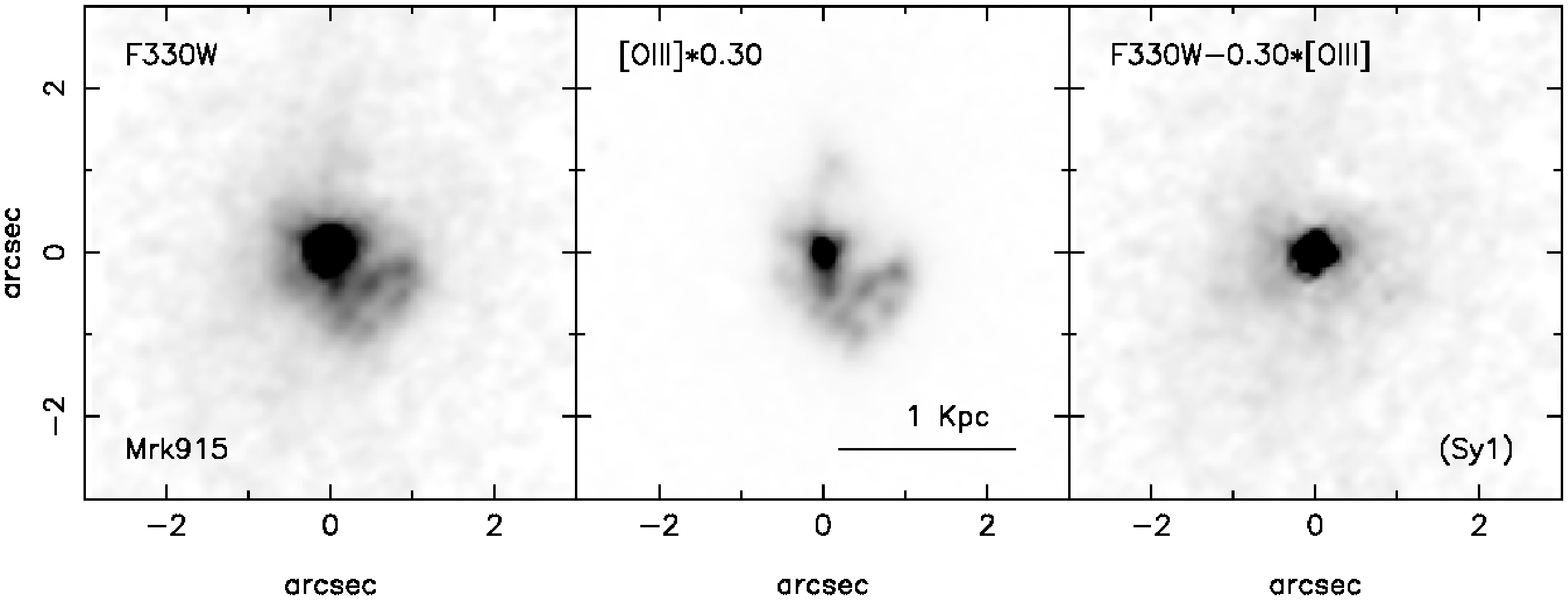}
\caption{The panels show sets of three images showing the same field of view, and plotted with the same intensity scaling. Left panel: F330W ACS image convolved with WF3 PSF; middle panel: [OIII] image scaled by the indicated factor (see discussion in text); right panel: residuals from the subtraction of the images in the middle panel from those in the left panel. North is to the top; East is to the left.}
\label{fig:all3_01}
  \end{minipage}
\end{figure*}                                           

\addtocounter{figure}{-1}

\begin{figure*}
  \centering
  \begin{minipage}{160mm}
\includegraphics[angle=-90,width=\textwidth]{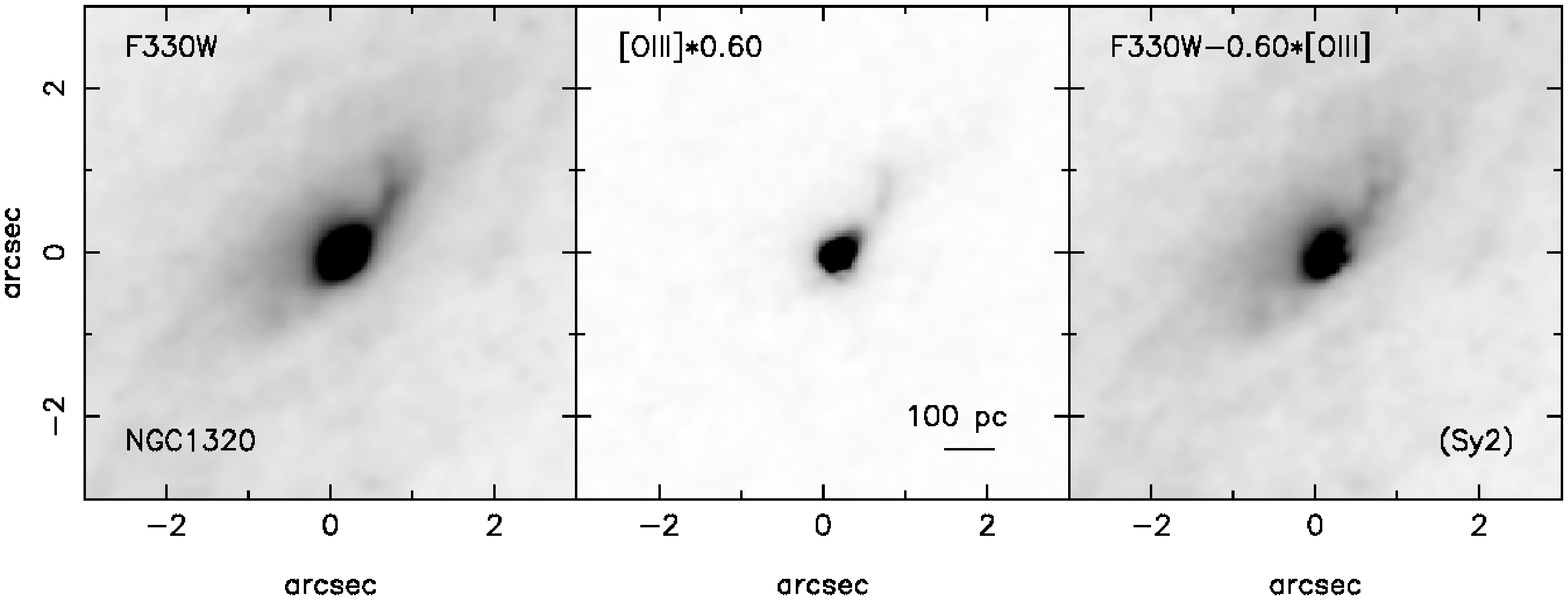}
\includegraphics[angle=-90,width=\textwidth]{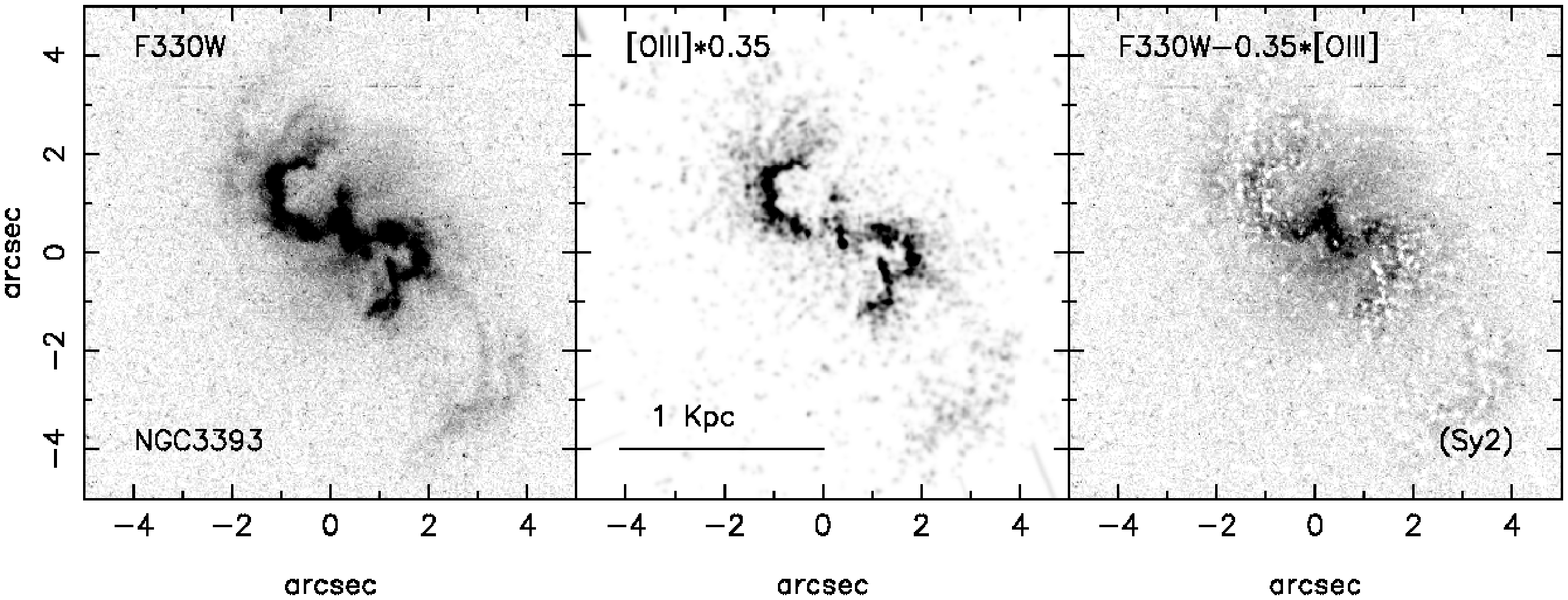}
\includegraphics[angle=-90,width=\textwidth]{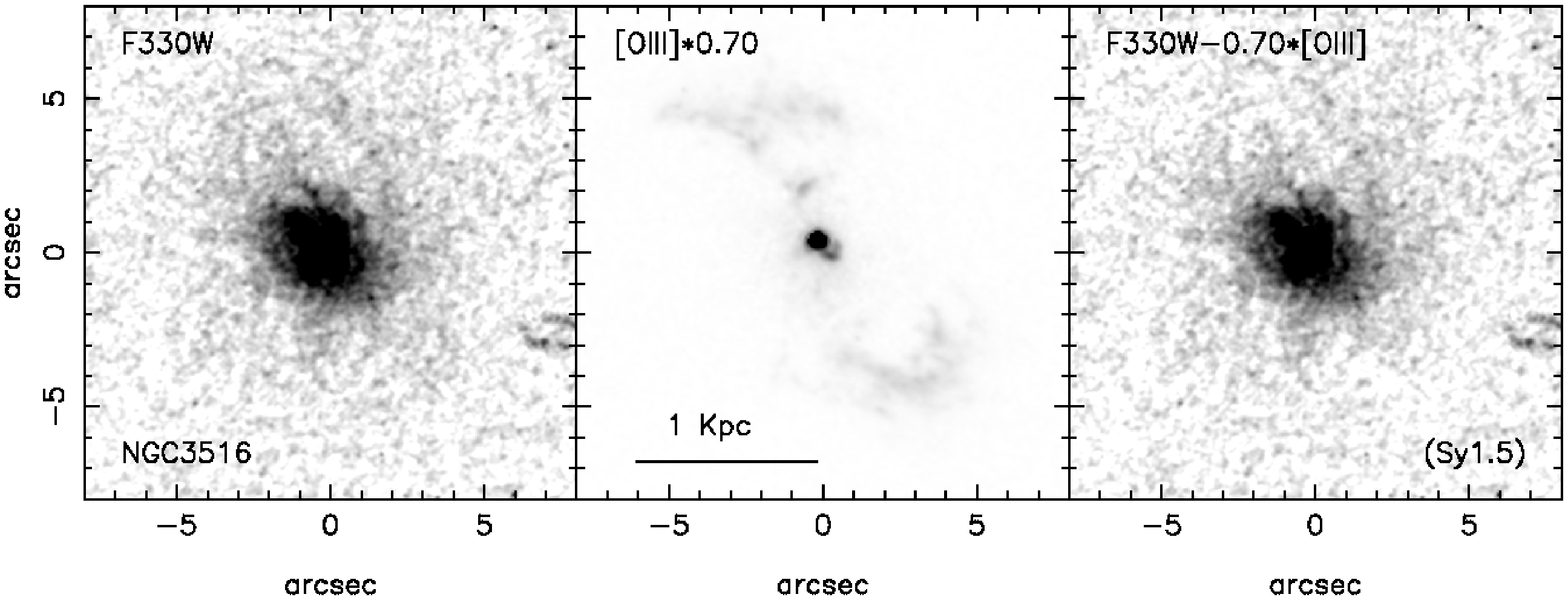}
\caption{Cont.}
\label{fig:otras-nev}
  \end{minipage}
\end{figure*} 

\addtocounter{figure}{-1}

\begin{figure*}
  \centering
  \begin{minipage}{160mm}
\includegraphics[angle=-90,width=\textwidth]{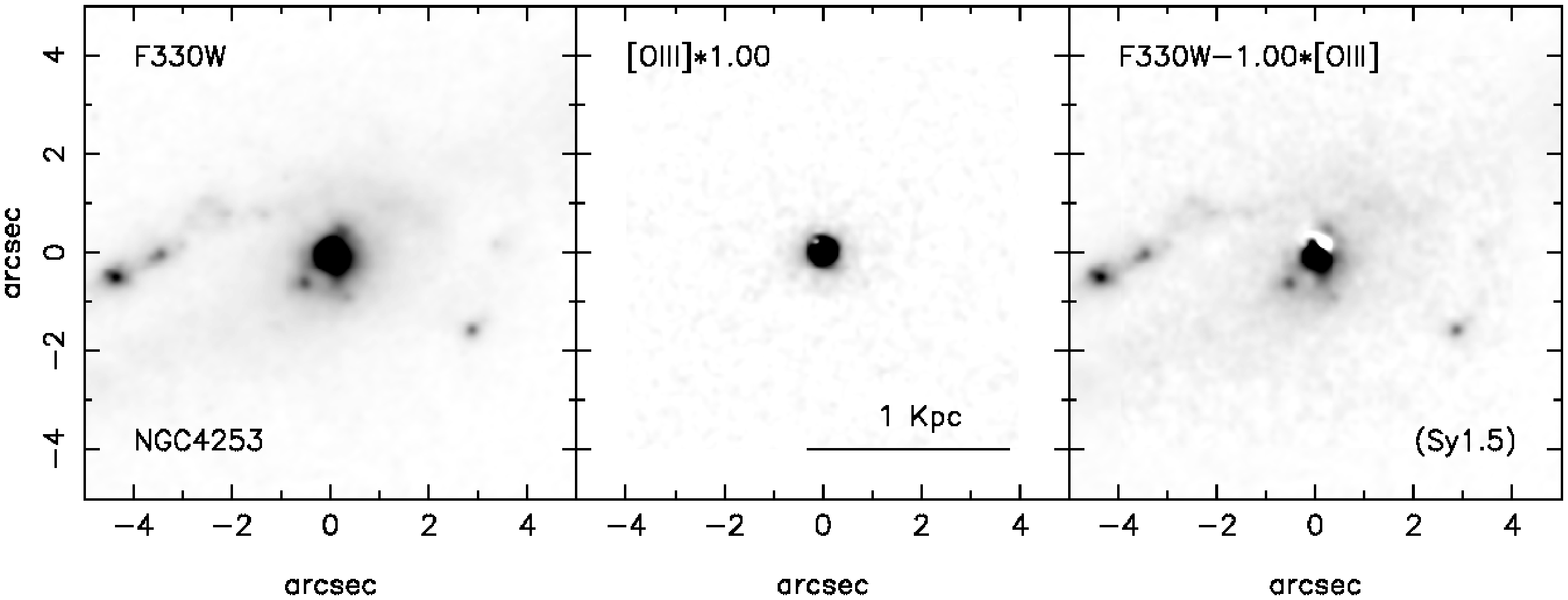}
\includegraphics[angle=-90,width=\textwidth]{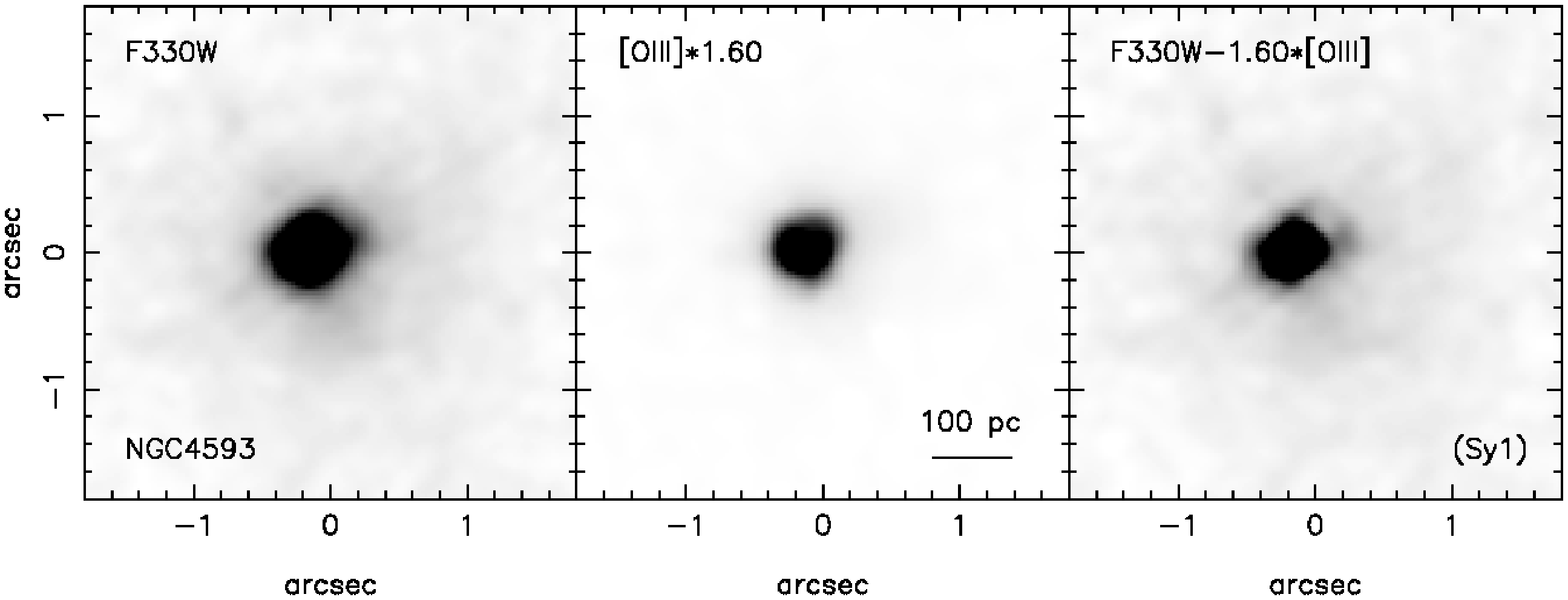}
\includegraphics[angle=-90,width=\textwidth]{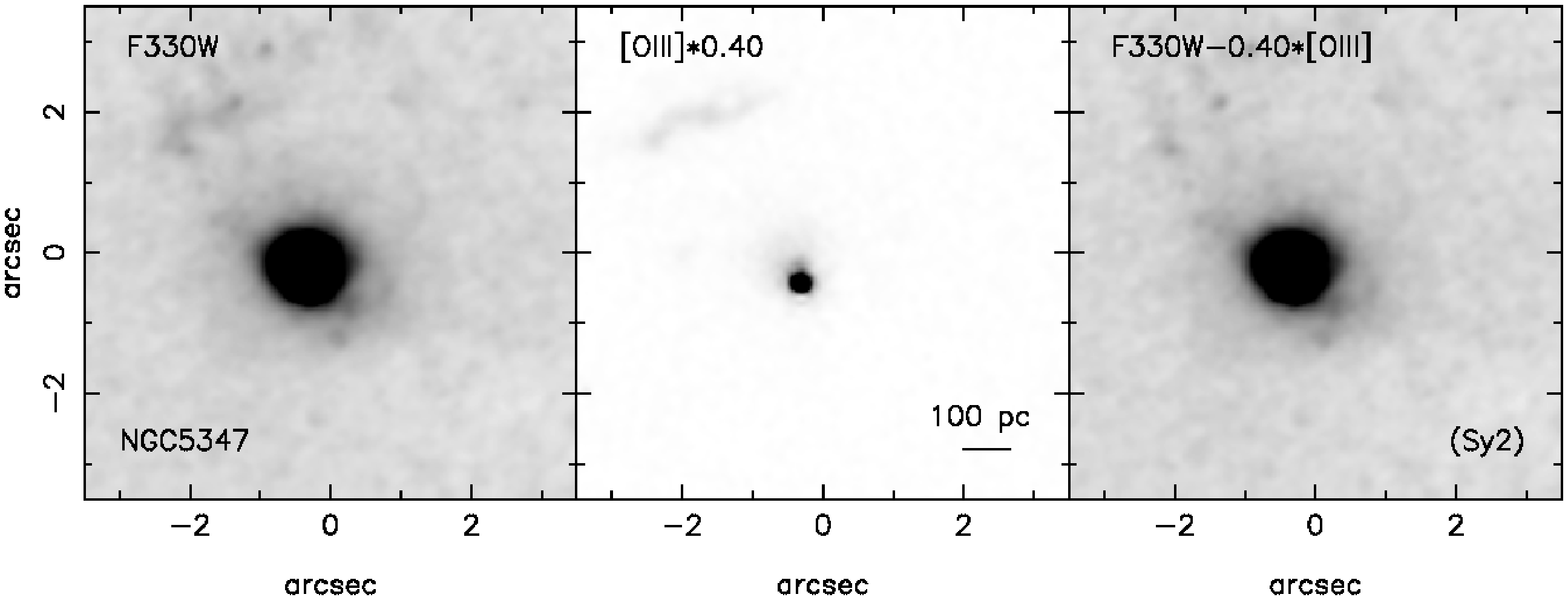}
\caption{Cont.}
  \end{minipage}
\end{figure*} 

\addtocounter{figure}{-1}

\begin{figure*}
  \centering
  \begin{minipage}{160mm}
\includegraphics[angle=-90,width=\textwidth]{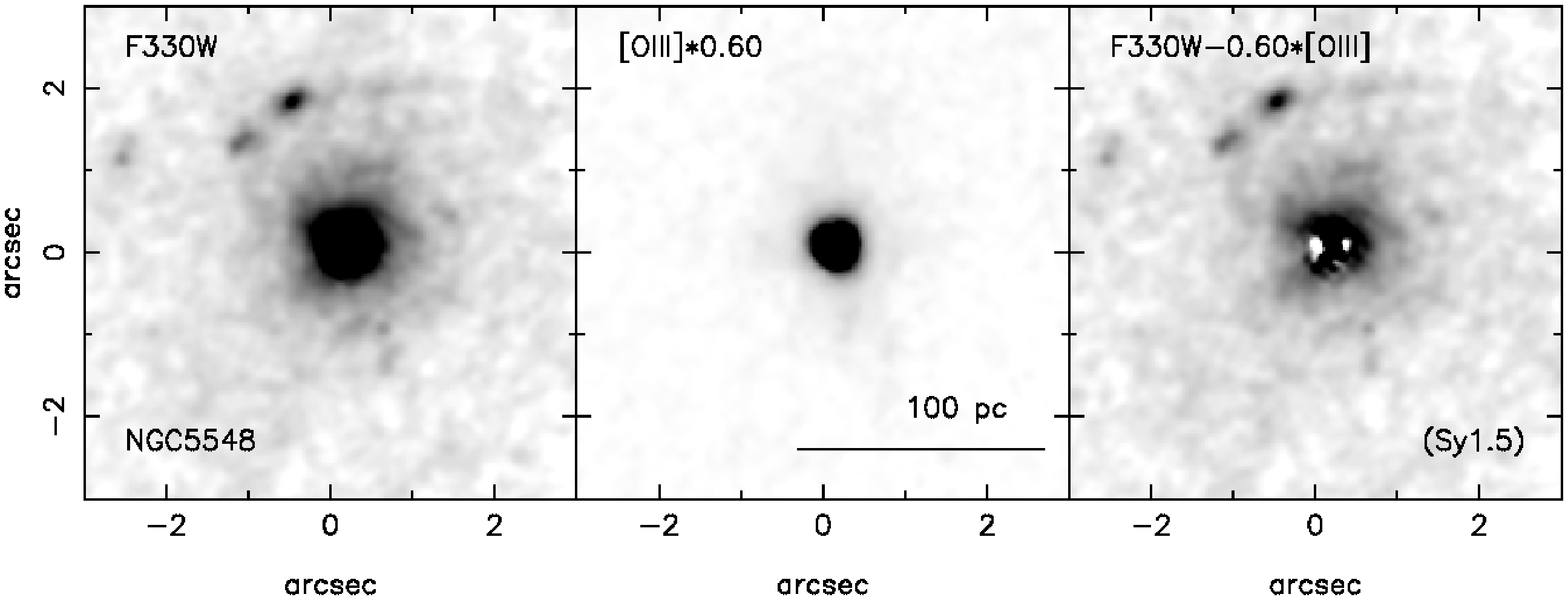}
\includegraphics[angle=-90,width=\textwidth]{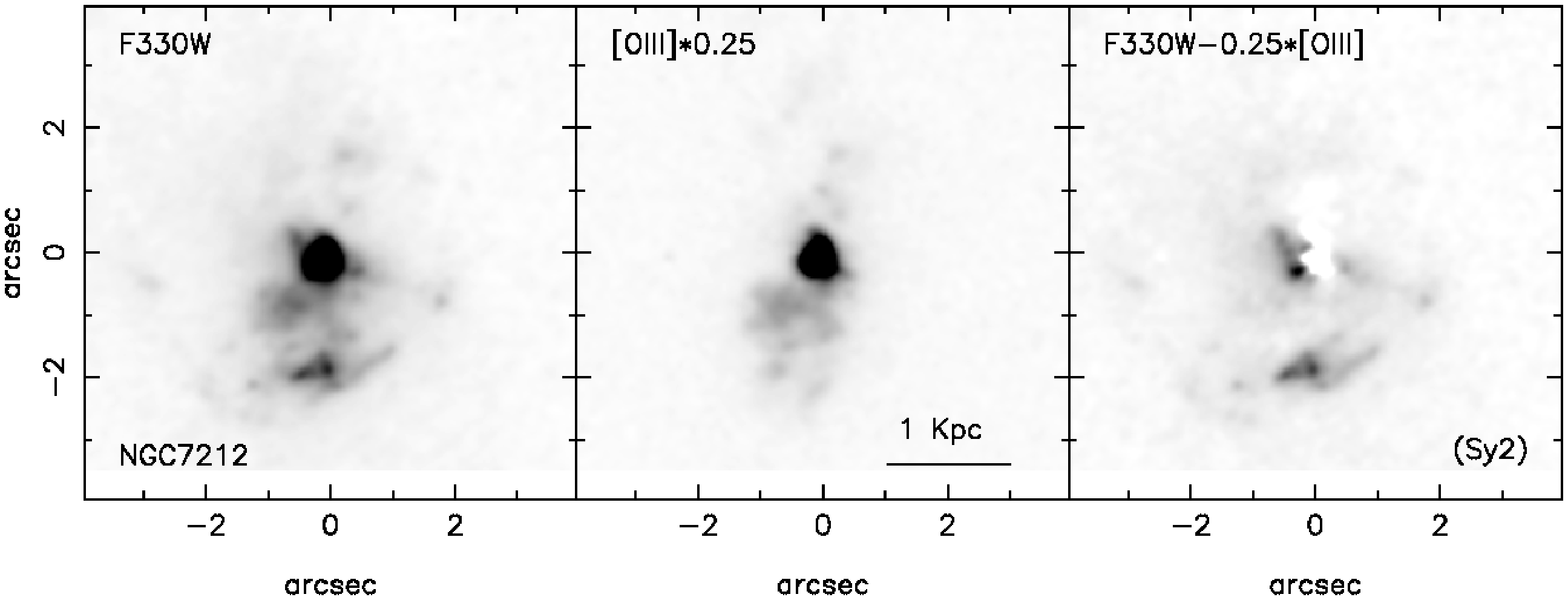}
\includegraphics[angle=-90,width=\textwidth]{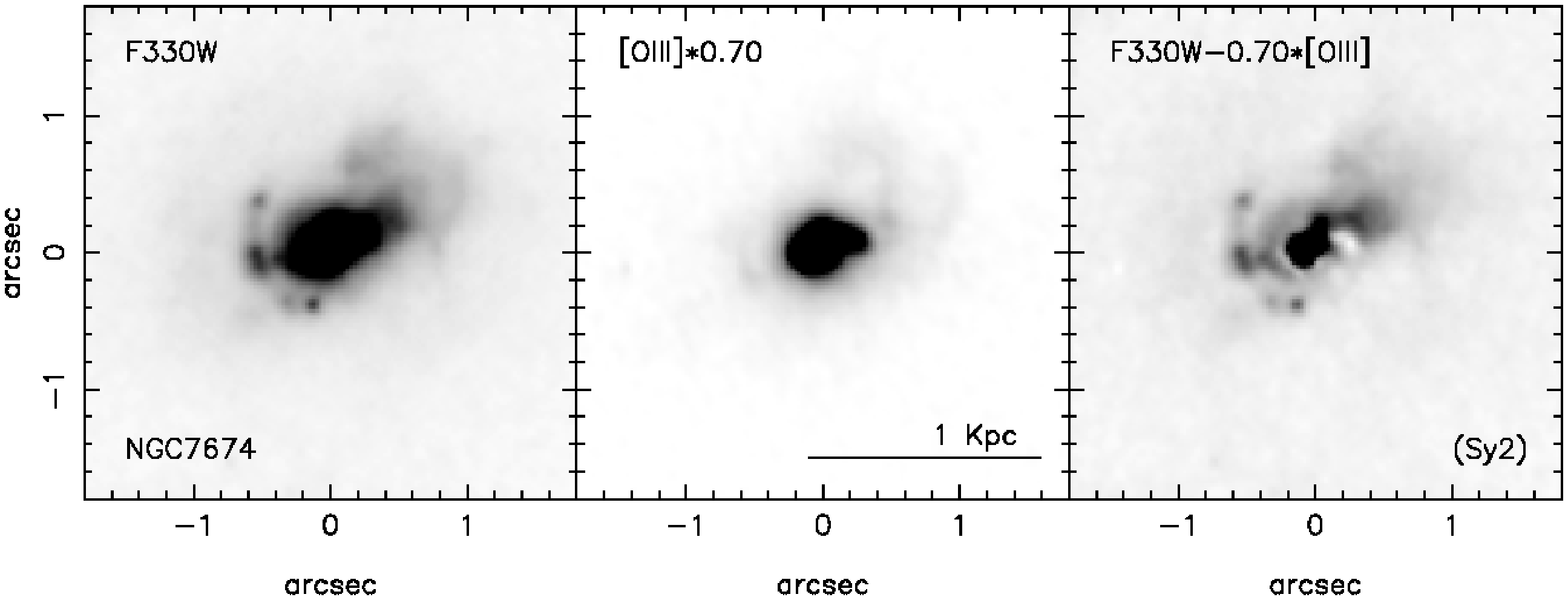}
\caption{Cont.}
  \end{minipage}
\end{figure*} 

\addtocounter{figure}{-1}

\begin{figure*}
  \centering
  \begin{minipage}{160mm}
\includegraphics[angle=-90,width=\textwidth]{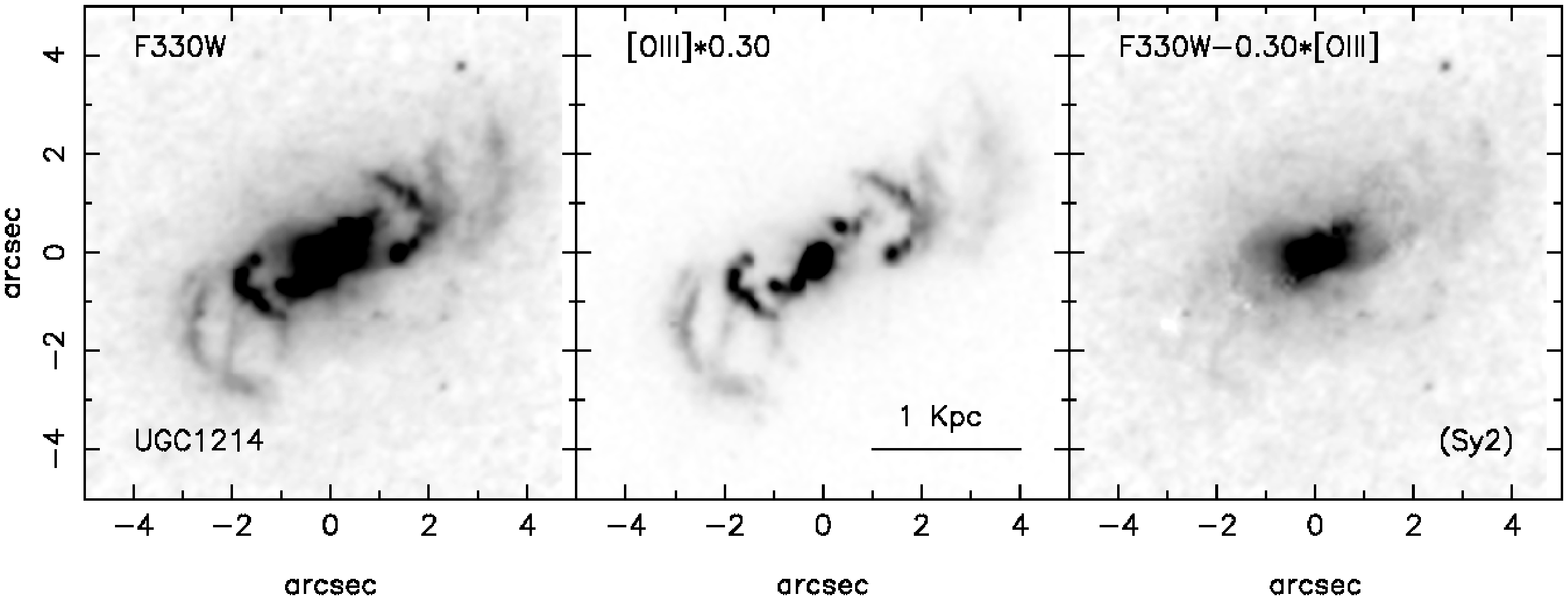}
\includegraphics[angle=-90,width=\textwidth]{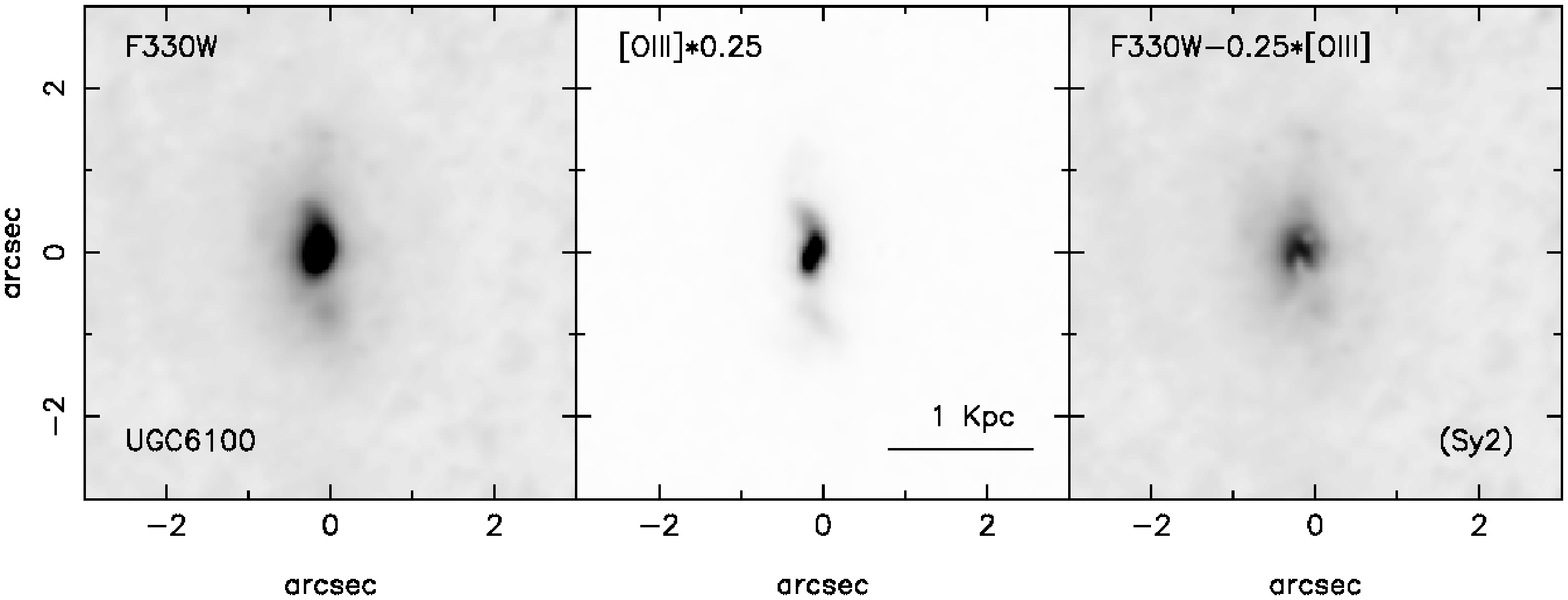}
\includegraphics[angle=-90,width=\textwidth]{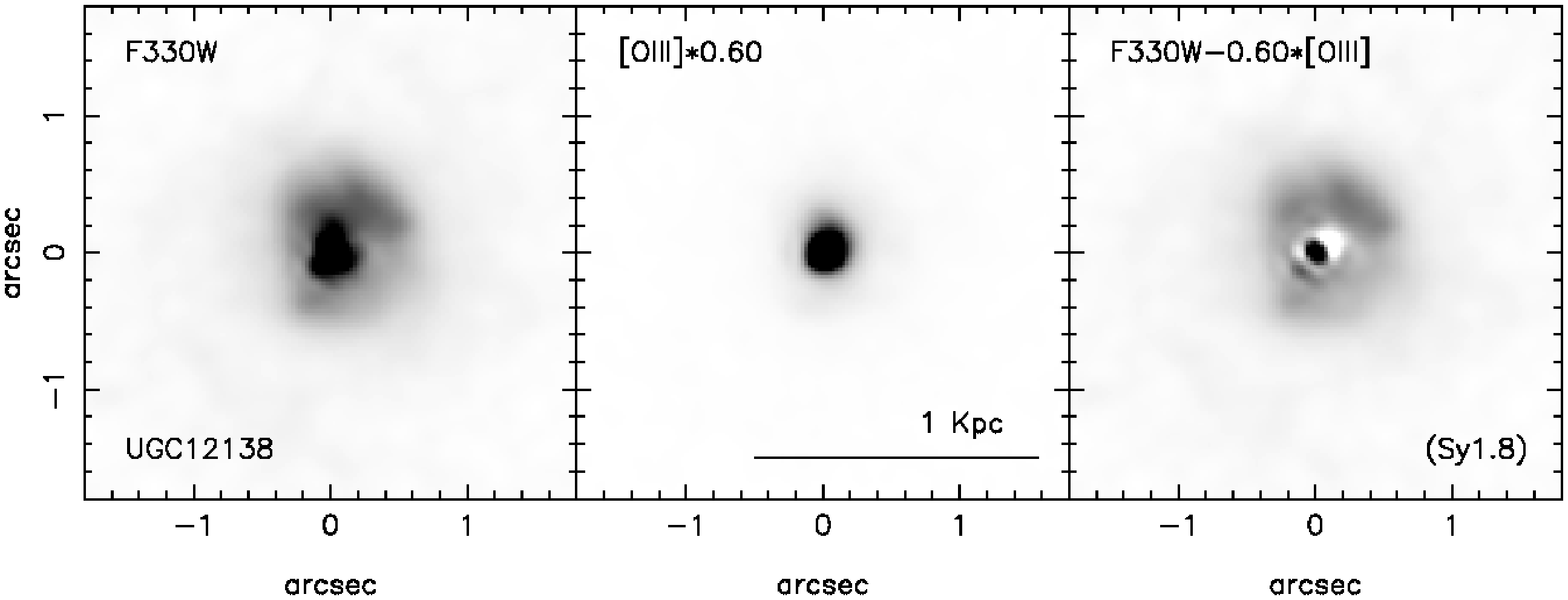}
\caption{Cont.}
  \end{minipage}
\end{figure*}



\subsection{Analysis: photometry of the residuals and other emitting mechanisms}
\label{sec:Phot}

After the subtraction of the ionised gas component (using the synthetic [NeV] + NC images), the objects showed still residual near-UV emission. In this section, we use some of these images in order to investigate the nature of the residual light. 

F330W images were degraded to the resolution of WFPC2, after removing the gas contribution with the factor {\bf f} estimated in Section \ref{sec:UV-oiii}. The addition of WFPC2 F547M and NICMOS F160W images, allowed us to study two colours. However, many of the objects show a bright compact nucleus, which may affect the study of the extended emission in the very inner regions of the galaxy. In addition, at the F160W band there could be a contribution of the torus emission, difficult to account for. Therefore, we will avoid studying the central region when the image is dominated by a strong nuclear point-like source contribution. 

Aperture photometry was performed with the IRAF task `phot'. Depending on the morphology of each object in particular, we calculated the colour radial structure, or integrated the light in certain apertures or rings. Comparing the results with synthetic photometry (see bellow) we can derive conclusions about the emitting mechanisms for some objects. We present the results for each galaxy in the next section.

\subsubsection{Synthetic photometry}
\label{sec:synPhot}
In order to compare our photometric measurements with the expected colours of a certain stellar population, we have obtained the synthetic colours (F547M--F330W and F160W--F547M) using the Spectral Energy Distribution (SED) of several sets of Single Stellar Population (SSP) models. Synthetic photometry was computed by multiplying each SED by the filter transmission curve plus detector response, and then comparing that to the result of doing the same with a reference spectrum, what defines the magnitude scale used. We tested the accuracy of our measurements with our own software using the SYNPHOT IRAF package. We chose the STMAG magnitude system, which uses a reference spectrum that has a constant flux density per wavelength interval. Colour indexes were then calculated from the magnitudes at F330W, F547M and F160W ($\sim$\,{\it U}, {\it V} and {\it H}) bands.

We have used model SED obtained using two different codes for spectral synthesis modelling: Starburst99 code (hereafter SB99; Leitherer et al.~1999, or V\'azquez \& Leitherer 2005) and the models of Bruzual \& Charlot (2003, hereafter BC03). The SSP models were obtained using:\\
{\bf a)} SB99 with the Padova 1994 and Geneva 1994 tracks, assuming a Salpeter IMF with mass limits between 0.1 and 100~M$_\odot$. We used the Lejeune stellar atmospheres with solar metallicity. We covered a range of ages from 1~Ma to 10~Ga, with a non-uniform age interval.\\
{\bf b)} BC03 also with the Padova and Geneva 1994 tracks, 
 Salpeter IMF and solar metallicity (see Bruzual \& Charlot 2003 for more details), with the same age range than above.

\begin{figure}
\includegraphics[angle=0,width=0.5\textwidth]{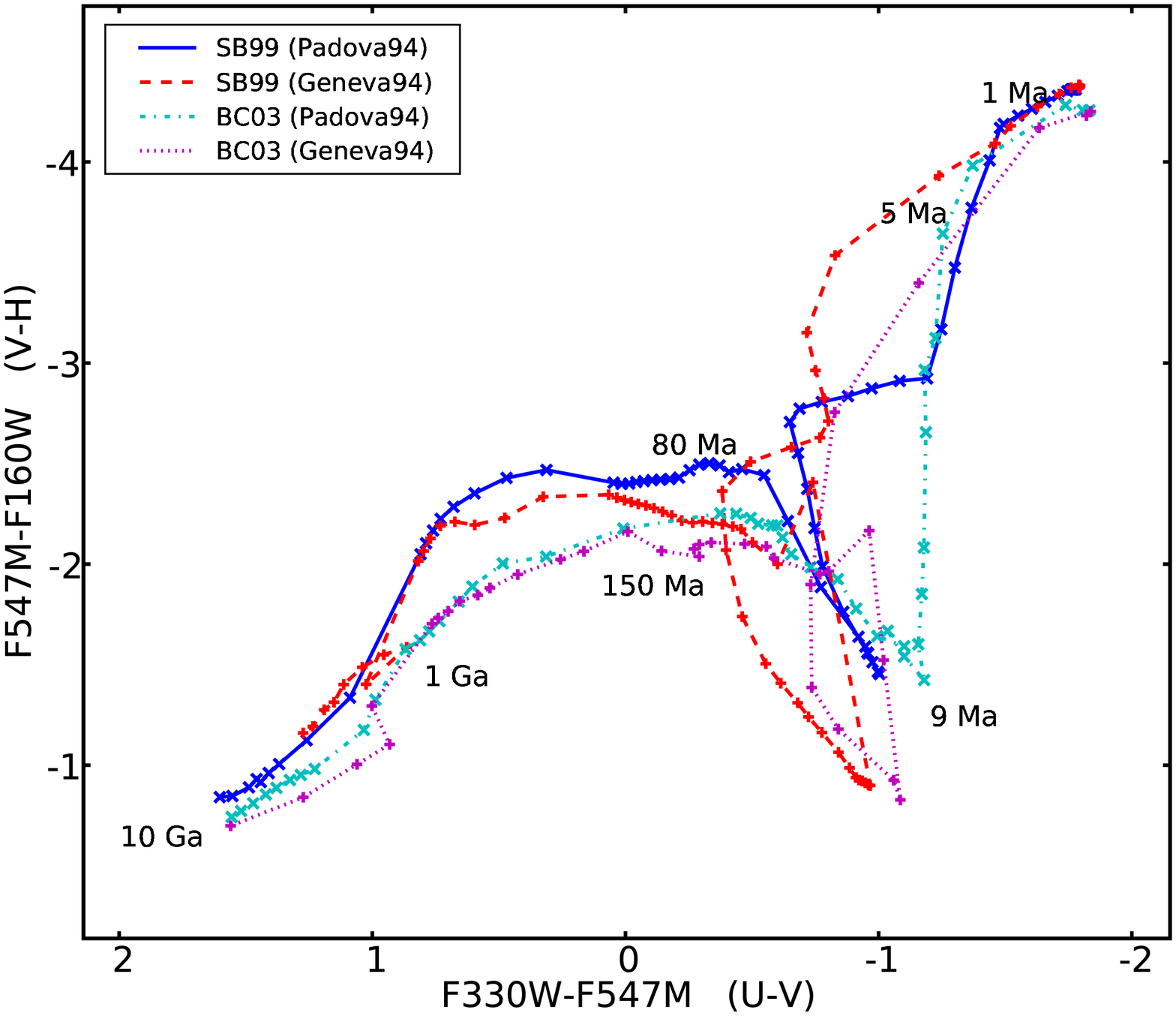}
\caption{Evolution of the studied colours with age of a Single Stellar Population model in a (F330W--F547M)~vs~(F547M--F160W) diagram. The graph shows the comparison of Starburst99 (SB99) and Bruzual \& Charlot 2003 (BC03) models for both, Padova 1994 and Geneva 1994 tracks. All the models have a standard Salpeter IMF function. In blue full line, SB99 models with Padova tracks; in cyan dotted-dashed line, BC03 with Padova tracks; in red dashed line, SB99 with Geneva tracks; magenta dotted line stands for BC03 models with Geneva tracks.}
\label{fig:UV-VH_comp}
\end{figure}

We have chosen these two codes because they are well recognised and frequently used throughout the literature. Both models allow for the use of two different well-known stellar evolutionary tracks: Padova and Geneva tracks. Stellar tracks are critical in order to determine the result of the models. Synthesis models have many additional ingredients affecting the output spectra, such as the Initial Mass Function (IMF) or the stellar atmospheres used for different evolutionary faces. Even using the same ingredients, different synthesis codes may output different results, due to the numerical implementation, as for example, different methods of interpolation within the evolutionary tracks. Thus, it is not surprising to find discrepancies in the photometry using output spectra from different codes.
 This can be noted in Fig.~\ref{fig:UV-VH_comp}, where we plot the time evolution of F330W--F547M and F547M--F160W colours of SSP models from SB99 and BC03. Both sets of models use a standard Salpeter IMF with mass limits between 0.1 and 100~M$_\odot$. The discrepancies are due to the effects commented above. As explained in V\'azquez \& Leitherer (2005), Geneva tracks fit better the young stellar populations, while Padova tracks are optimised for older stellar populations.

\begin{figure}
\includegraphics[angle=0,width=0.5\textwidth]{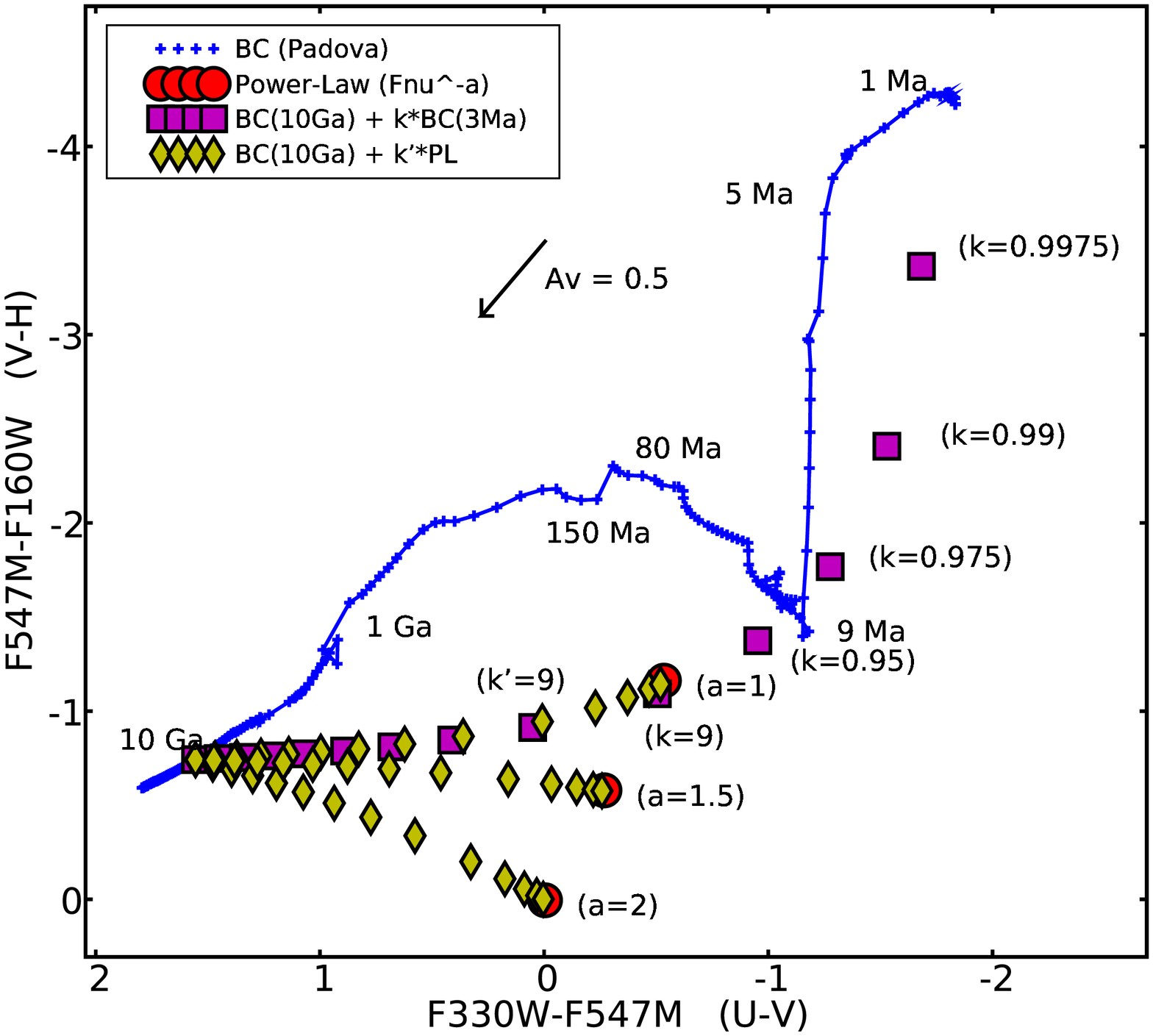}
\caption{Same colour-colour diagram than Fig.~\ref{fig:UV-VH_comp}, showing the location of some power-law spectra, and two sequences of a 10~Ga old stellar population added to: (1) a young population of 3~Ma (magenta squares), and (2) a power-law $f_\nu \propto \nu^{-a}$ with $a$=$1, 1.5, 2$ (green diamonds). For reference, we have also plotted the reddening vector, and the evolution of a SSP model of Bruzual \& Charlot (2003), with Padova tracks and Salpeter IMF. The scaling factors k and k' are related to the fraction q, described in text, as $q=k/(1+k)$.}
\label{fig:degPL9Ma}
\end{figure}

We have also calculated the expected location of AGN light in the (F330W--F547M)~vs~(F547M--F160W) diagram. We performed synthetic photometry through the given filters of power-law spectra with indexes (in flux density per unit frequency, defined as $f_\nu \propto \nu^{-a}$) 1, 1.5, and 2. This is the expected range for the index of the spectrum of an AGN nucleus at optical-near-UV wavelength (Osterbrock 1989), being more common the smaller values. The results are redder colours than those of the youngest stellar populations. 
The presence of scattering by dust might make the colours bluer, although not necessarily if dust is distributed in a clumpy medium (see Vernet et al.~2001; note that free electrons produce also a grey scatter, wavelength independent). However, the scattered light, as well as any young stellar population, is most likely found diluted in an old bulge population, what would shift the colours to redder values. Different mixture fractions with the red population may lead to confusion between the contribution of a diffuse young stellar population and the scattered AGN light. Reddening processes may affect the colours according to the same trend. 
In Fig.~\ref{fig:degPL9Ma} we illustrate this degeneracy, by comparing the synthetic photometry of the power-laws with that obtained from a sequence of SSP Bruzual \& Charlot (2003) models. We also include in this figure the photometry of two series of spectra: (1) the combination of a young 3\,Ma old stellar population to a 10\,Ga old one and (2) the combination of a power-law with exponents 1, 1.5, and 2, to the same \mbox{10\,Ga} old stellar population. The normalisation is such that, at the average wavelength of the blue filter (336.72\,nm, calculated with SYNPHOT package), the contribution of the bluer spectrum to the total flux is a factor \emph{q}. We used q values in the range 0--0.9 with increments of 0.1 (10 per cent of the total contribution), plus  0.95, 0.975, 0.99, and 0.9975. In Fig.~\ref{fig:degPL9Ma} it is appreciated how a small contribution (note: small in flux, not in mass) of an old stellar population shifts the colours of the blue spectra to positions where it can be confused with reddened older stellar populations. In addition, the 3\,Ma and the power-law sequences follow a similar path in the diagram. 
 We can conclude that, in general, it will be very difficult to disentangle the two mechanisms with the available filters. However, the colour analysis can give us important clues for particular objects.

\section{RESULT OF THE ANALYSIS FOR EACH INDIVIDUAL OBJECT}
\label{sec:results}

In this section we give details of the analysis procedure followed for each object. The particular morphology of each nucleus, and the availability and quality of the data, put some constrains on the kind of analysis we can perform in each case. For example, in some cases it is convenient to perform a radial colour profile, as illustrated in Fig.~\ref{fig:U-V}, while other times we just carry out aperture photometry in one or more regions. We have attempted to extract the most from the images of each galaxy. The measurements discussed below, and plotted in Fig.~\ref{fig:result}, are corrected for galactic extinction only, using the law of Cardelli et al.~(1989), and the E(B-V) values from NED\footnote{The NASA/IPAC Extragalactic Database, which is operated by the Jet Propulsion Laboratory, California Institute of Technology.}. The results are also summarised in a tabular form in Table~\ref{tab:res}, in which we include the relative contribution of each component when possible, as well as a summary of the photometry of the objects after removing the Sy nucleus and the contribution of the ionised gas. Note that the contribution of ionised gas is given respect to the F330W flux once the point-like nucleus is subtracted. In this way it is possible to do a direct comparison between objects with and without nucleus. In order to present the results in a more homogeneous way, we have chosen to give the fluxes and fractions in a physical region of 1\,kpc radius around the centre of each object. 
Along this section we commonly write {\it U--V} and {\it V--H} when referring to F330W--F547M and F547M--F160W colours.

\subsection{IC\,5063 (Sy2)} 
Most of the emission to the southeast is removed with {\bf f}=0.20, but in order to get rid of the northern filamentary structure, the scaling factor has to be increased to 0.25-0.30. Some negative residuals appear in the subtracted image (top right panel of Fig.~\ref{fig:all3_01}). Rather than a change in the ionisation parameter, we suspect that these residuals may be due to dust obscuration, affecting much more the near-UV image than the [OIII] one. This object is heavily obscured by dust, which can explain the axial asymmetry of the emission. This is better appreciated in the F330W filter, as the bulk of the emission is concentrated to the southwest of the line northwest--southeast. The largest negative residuals after the subtraction process are found in the northern border of this axis, where we expect the obscuration to be larger. After the subtraction there is much extended emission to the southwest and a couple of clumps left. We have performed aperture photometry in three regions of radius 0.75 arcsec next to the nucleus (1--2 arcsec distance to the southwest of the peak of the near-UV emission). The photometry yields the colours {\it U--V}$\sim$1.7, {\it V--H}$\sim$-0.6. These are compatible either with a very old stellar population (older than 10\,Ga), or with an old obscured population. 
The bright clump which lies $\sim$3 arcsec to the northwest shows a somewhat bluer {\it U--V} colour (1.07). This may indicate the contribution of localised star formation, or  scattered light from the AGN in a denser region of the ionisation cone.
 From the literature data and the estimated {\bf f} value we infer that the nebular continuum alone can account for the light coming from the filamentary structure (23 per cent of the near-UV emission in the inner kpc), while most of the remaining light is stellar continuum from the bulge ($\sim$77 per cent of the total emission in the inner kpc).

\subsection{Mrk\,6 (Sy1.5)} 
This is a Sy1 galaxy with a bright nucleus. After the best nuclear subtraction some positive residuals are left, which are seen as a compact, though slightly extended, source in the convolved image. The factor {\bf f} is relatively small for this object. A value of 0.2 accounts for the brightest regions and blobs to the north and south of the nucleus. At the same time it does not leave strong negative residuals, as f=0.3 does. Once the bright Seyfert nucleus is removed, the ionised gas emission account for an important part of the remaining extended emission ($\sim$60 per cent). Some nuclear emission remains, as well as an extended component. In addition, there seems to be a narrow diffuse filament extending towards northeast, and perpendicular to it, two small arcs in a symmetric structure both sides of the nucleus. This structure may be related to the radio emission. Kharb et al.~(2006) report a double-bubble structure in 6 cm VLA radio-maps, and one of the directions with strongest emission matches the orientation of the near-UV filament (see fig.~1 of their work).
In NICMOS F160W, and WFPC2 F547M images, the galaxy appears completely dominated by the bright nucleus, so we will not perform a further photometric analysis of the residuals. 

\subsection{Mrk\,915 (Sy1)} 
The best value of {\bf f} is around 0.30, with which most of the flattened-O shaped emission to the southwest is subtracted. There remains however a strong extended emission, that is unlikely to be due to ionised gas.
This galaxy is a Sy 1 with a strong nuclear component. As in the case of Mrk\,6, the best nuclear subtraction still leaves some positive residuals in the centre which, after convolving with the PSF of WFPC2, resemble a compact nuclear source. 
One could argue that a too conservative PSF subtraction may mimic that effect, although this is unlikely, as we stopped the subtraction when negative residuals began to appear. In order to check how much of the extended emission would be due to light from the extended wings of the PSF in case of a poor PSF subtraction, we subtract to the near-UV image a PSF (convolution of F330W PSF with one of the WF3 at 500.7~nm) of flux \mbox{3e-13 erg/cm$^{2}$/s}. This subtracts most of the light in the inner 0.6 arcsec. From aperture photometry of this image and the scaled [OIII] subtracted one, we can set a very conservative upper limit, estimating that at 0.75 arcsec from the nucleus, no more than 30 per cent of the light can be due to a bad subtraction of the central source.
Therefore, there is another mechanism beyond ionised gas emission and contamination by a poor PSF subtraction, that is producing the extended UV halo and compact emission. WFPC2 and, to a greater extent, NICMOS images, are dominated by a strong nuclear emission, so any further analysis would be dominated by the nuclear subtraction uncertainties.

\subsection{NGC\,1320 (Sy2)} 
\begin{figure*}
\centering
\begin{minipage}{160mm}
%
\includegraphics[angle=0,width=0.5\textwidth]{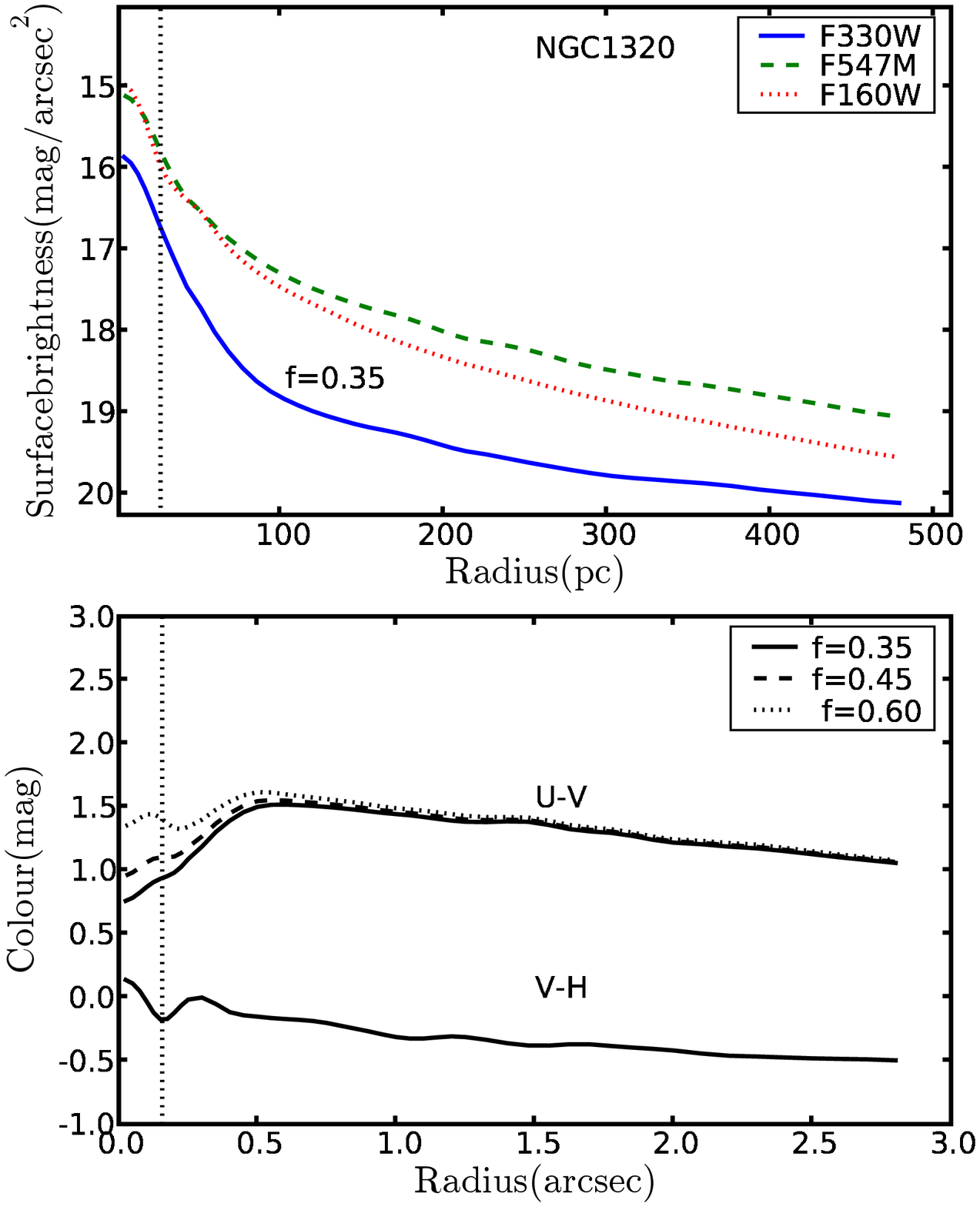}
\hfil
\includegraphics[angle=0,width=0.5\textwidth]{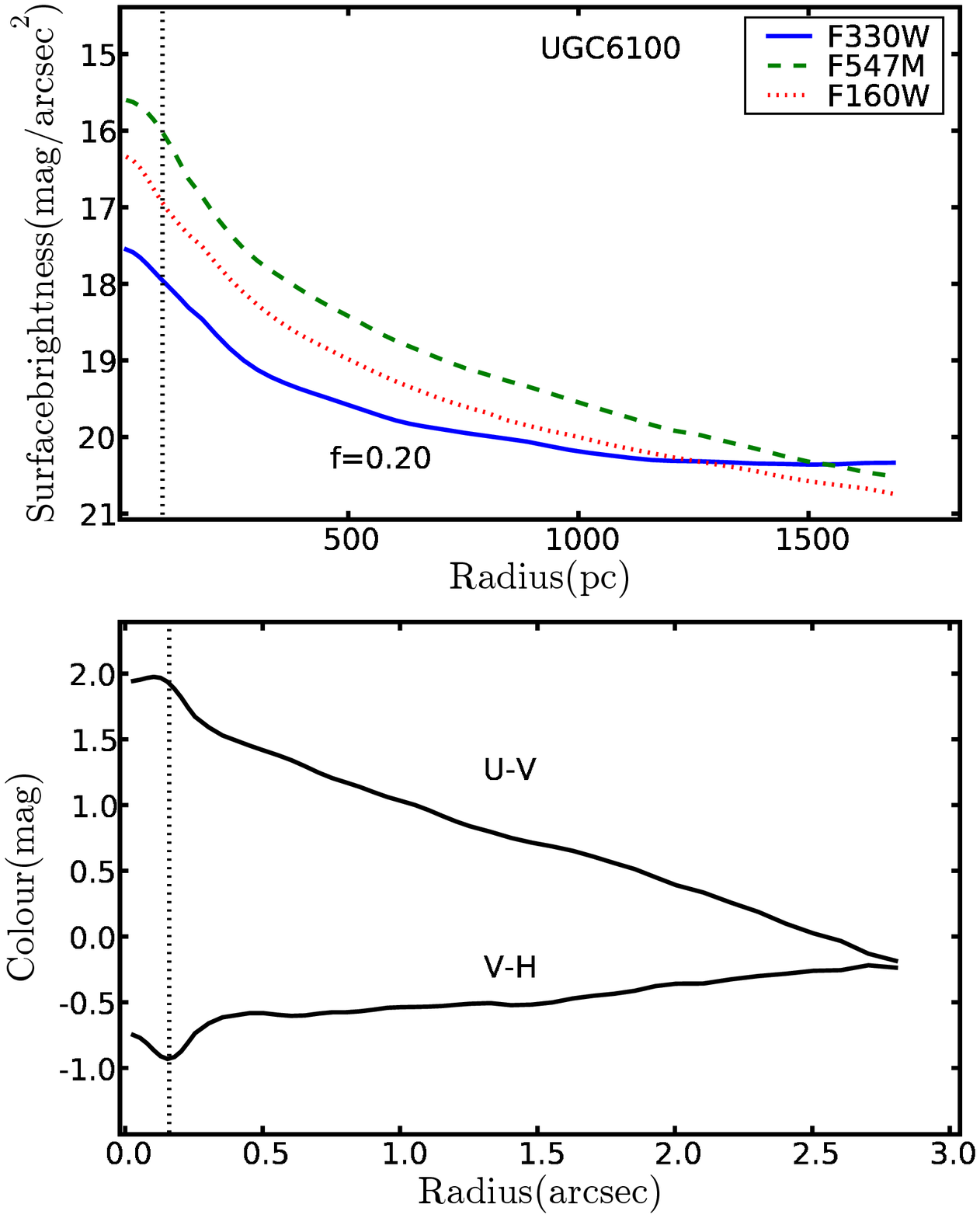}
\caption{A photometric comparison of two Sy2 with a strong central near-UV emission, which cannot be accounted for by emission from ionised gas. In the upper panels we plot the surface brightness profiles for the filters F330W (blue full line), F547M (green dashed line), and F160W (red dotted line), which were obtained with circular apertures. Bellow we show the radial dependence of the colours F330W--F547M and F547M--F160W. The FWHM of the PSF is marked with a vertical dotted line. A central component of different nature for each object is found in the inner 100\,pc (see discussion in text).}
\label{fig:U-V}
\end{minipage}
\end{figure*}

This galaxy is an interesting case. In order to try to remove the filament to northwest, we have used f~=~0.6, which is twice the average value obtained for other cases. Some structure is still seen, but applying a higher factor will introduce strong negative residuals next to the nucleus which are difficult to justify. The value of {\bf f} as high as 0.6 might be justified as possibly due to a smaller [OIII]/H$\beta$ ratio in this galaxy, but still it does not explain all the emission. Most likely, it has a regular {\bf f} ($\sim$0.3) and a further contribution in the central region. Considering that {\bf f} $\lesssim$0.6, photometry in the inner 500\,pc sets an upper limit of 15 per cent for the contribution of the nebular emission to the total near-UV light. 
We have carried out a circular aperture photometric study in the residual image. The profile shows a nearly constant U--V colour gradient of $\sim$~--0.12\,mag/100\,pc at radii larger than 100\,pc (compare with UGC\,6100), and a central blue drop, whose depth depends on the adopted {\bf f} factor. However, even with values as high as f=0.6, the inner drop is evident in the inner 100\,pc (see Fig.~\ref{fig:U-V}). 
This indicates that there is a contribution either from a young population in a circumnuclear disc, or from scattered AGN light. 
This result is confirmed measuring the colours in a 0.4 arcsec radius aperture around the nucleus. The colours ({\it U--V}=1.06, {\it V--H}=-0.25) point to the existence of an old bulge population with a significant bluer component. %
This conclusion does not change using neither slightly wider or narrower apertures, nor applying previously a different scaling factor to the ionised gas image (between 0.2 and 0.35). Cid Fernandes et al. (2001) found that $\sim$10 per cent of light at 486.1~nm is due to young/intermediate age stars and/or a power law component. However, spectropolarimetry observations did not find any polarised broad emission in this galaxy (Tran, 2001), so the likelihood of the blue
light being scattered light is very small, this emission is most likely due to star formation. 
It is interesting that we obtain such a red {\it V--H} colour for this nucleus. Further than 200~pc the colours are closer to the ones we obtain for other objects, but with a significant blue component. In the inner region the nuclear torus emission contributes to the light through the F160W filter. However, this emission is visible as a point-like source (Quillen et al.~2001). Between 100 and 200\,pc from the nucleus we have to invoke either a high reddening or another contribution to F160W, apart from stellar emission, that we are not taking into account.

\subsection{NGC\,3393 (Sy2)} 


The [OIII] image of this galaxy was obtained with WF/PC before COSTAR, so it had to be deconvolved. The resulting image is quite clumpy, an artifact of the deconvolution process. The artifact structures remains after convolving with the ACS PSF. So a detailed analysis, as we do with the other objects, will not be possible for this galaxy. In this case we compare the images directly, without convolving first with the PSF of the other band, taking into account that the resolution for the two images is quite similar.

An scaling factor of 0.20 removes the brightest part of the filaments. With a factor of 0.50, most of the structure is gone, although strong negative residuals are left. The best result is achieved with  {\bf f} $\sim$~0.30--0.35. This fits with the expected value of 0.35, calculated from the data in Table~\ref{tab:oiii_sample}. The range of {\bf f} values necessary to remove the different parts of the filaments might be due to the clumps mentioned above, as well as to a change in the ionisation parameter along the filaments. Most of the central structure in the F330W image does not seem to be due to ionised gas (remains after subtraction with {\bf f}=0.50). Through photometry in theinner kpc we estimate a contribution of the ionised gas of $\sim$33 per cent to the total near-UV light, with a higher uncertainty than for other objects, due to the artifacts of the reduction process. Because of this uncertainty we have not attempted a further photometric analysis, although the estimated value of {\bf f} supports the results from similar objects such as UGC\,1214 (see below).

\subsection{NGC\,3516 (Sy1.5)} 
After removing the Seyfert nucleus, there remains a diffuse emission around it, which shows no clumps or star clusters. For this object the morphology in the near-UV and [OIII] is very dissimilar. This is probably the best example in the whole sample in which [NeV] and NC contribute very little to the near-UV light. Simple aperture photometry shows that in the inner kiloparsec, the contribution of the ionised gas accounts for less than 10 per cent (considering a scaling factor of {\bf f}=0.6) of the diffuse near-UV light. Smith et al.~(2002) find no intrinsic polarisation in this galaxy, which makes the scattered light explanation unlikely, in spite of earlier works showing a high polarisation for the same object (Martel, 1996). In order to understand the origin of this light we perform an aperture photometry analysis of the three bands. Curiously, the F547M image does not show a very bright nucleus, as the others, so the usual data reduction and resolution-matching process is enough for this band. On the contrary, the NICMOS image shows a bright point-like source, although in this case it doesn't preclude the study of the also bright bulge surrounding it. After the standard calibration process, we remove the near-IR nucleus using the same procedure explained for F330W images, with a PSF created with the TinyTim software. We avoid using the inner 0.4 arcsec of the image, as the central pixels are affected by the residuals of the subtraction of the PSF. The analysis yields fairly constant colours ({\it U--V}$\sim$1.4, {\it V--H}$\sim$-0.6) in the whole studied range (from 0.4 arcsec up to 5 arcsec), which are the expected colours for an old stellar population. We thus conclude that the bright underlying bulge is responsible in this case for most, if not all, the extended emission in the UV.

\subsection{NGC\,4253 (Sy1.5)} 
After the subtraction of the PSF from the near-UV image of this Sy1.5 galaxy, there still remains a central bright compact source and several compact knots resembling massive star clusters or star-forming regions. The [OIII] emission in this galaxy is very compact. A literature based estimate for {\bf f} (from Table~\ref{tab:oiii_sample}) yields 0.41. However, applying a scaling factor as high as {\bf f}=1 still leaves most of the nuclear emission, but negative asymmetric residuals begin to appear near it. Therefore, gas emission cannot account for most of the light in the near-UV. An scaling factor of 0.6 yields an upper limit of 16 per cent for the contribution of nebular emission to the inner kpc near-UV light. In the filter F547M, the total exposure time is 20 seconds, so only the bright nucleus can be seen with enough signal to noise, and no further analysis was attempted.

\subsection{NGC\,4593 (Sy1)} 
This is a Sy1 with a very bright nucleus. After the best nuclear subtraction there remain some residuals at the nucleus and a diffuse light component, but the emission is quite compact overall. The [OIII] emission is also very compact and mainly limited to the inner 1 arcsec (less than 200\,pc).
 Considering a scaling factor of 0.6 we can set an upper limit to the nebular continuum and [NeV] contribution of less than 8 per cent of the total flux inside the inner 350\,pc, limited to the region in which there is enough signal-to-noise for the [OIII] emission. As [OIII] emission appears to be more compact than the near-UV, this probably constitutes a strong upper limit. In order to study the compact light emission we would have to remove the PSF from the three bands, which yields too many uncertainties to conclude anything further about the inner region.

\subsection{NGC\,5347 (Sy2)} 
This galaxy has two components in the [OIII] image: a compact central emission, plus a narrow arc $\sim$3 arcsec to the north-east. The near-UV image is somewhat similar, although the compact component is much brighter than the arc, relative to the [OIII] image. Although a literature based estimate of {\bf f} yields a value of 0.58, the arc is subtracted with {\bf f}=0.4, while the central emission persists even with very high {\bf f} values. We have integrated the flux within the inner 620\,pc, finding that the contribution of the ionised gas is probably less than 4 per cent of the flux in F([F330W]). This implies that neither the nebular continuum nor the [NeV] can account for the extended emission near the centre. In the two bluer bands (F330W and F547M) this emission is asymmetric, with the shape of a cone, opening towards the north-northeast. On the contrary, in the F160W filter (dominated by the bulge emission and the central AGN) the emission is quite symmetric. To avoid an azimuthal averaging, in this kind of object we prefer to perform aperture photometry in certain regions. We chose two small apertures of 0.25 arcsec radius to the north and northeast, including the extended emission but avoiding the nucleus. For the U band we used the F330W image minus the [OIII] contribution scaled with a factor {\bf f}=0.4. Using a literature based {\bf f}=0.6 yields a difference of only 0.04 magnitudes in the regions studied. In the NICMOS image we had to perform the subtraction of a PSF in order to eliminate the contribution of the nuclear source, as the apertures were only $\sim$0.5 arcsec far from the centre. The U--V colours obtained ({\it U--V}$\sim$0.9, {\it V--H}$\sim$-0.9) are around 0.5 magnitudes bluer than the Padova tracks (recommended for old stellar populations) for a population of 3~Ga. This implies that there is a contribution of a bluer component that may represent 10--30 per cent of the near-UV light (deduced from Fig.~\ref{fig:degPL9Ma}). Due to this low percentage, we think that the asymmetry of the emission in the bluer bands might be caused by dust obscuration, what seems obvious to the south of the nucleus also in the F547M image. Scattered light from the AGN could produce this cone shape naturally, but the contribution of the blue component to the total light seems too low to explain the morphology by this mechanism only.

\subsection{NGC\,5548 (Sy1.5)} 
As in the other Sy1 described above, this galaxy shows a compact near-UV emission, once the nuclear contribution is subtracted. Data in Table~\ref{tab:oiii_sample} suggest a high {\bf f} value for this object (0.67), although using such high scaling leaves strong negative residuals and still does not explain the bulk of the emission. From this we estimate that less than 25 per cent of the flux can be attributed to the ionised gas. 
Apart from the central light excess there is an extended surrounding emission, and a star-forming arc at $\sim$2 arcsec to the north-east. These features are not visible in the F547M image, due to its short exposure time (only 24 seconds). With only one colour, and the extremely bright nucleus dominating the near-IR, there is not much room for a further analysis of the residuals.

\subsection{NGC\,7212 (Sy2)} 
The diffuse emission region to the south of the nucleus is removed with {\bf f}=0.25. However this leaves negative residuals in the nuclear region and next to it to the north. It is expected that those regions suffer from a heavy absorption, as the usual bi-conical structure is shown here as a single cone towards the south. The scale factor needed to remove the small northern emitting zones is just 0.15. As explained in Section \ref{sec:NC}, this can be accounted for with an internal extinction of $A_V$=~1. 
The subtraction of the ionised gas emission by this method, allows us to identify the bright clumps to the south as star-forming regions, although some contribution by scattered light from the AGN might be expected near the nucleus (see Section~\ref{sec:scat}). Unfortunately, this interesting object lacks an F547M image, as well as any NICMOS exposure. 

\subsection{NGC\,7674 (Sy2)}
This galaxy is imaged with the PC, so the [OIII] image has a better resolution than for the other objects. The galaxy is a very blue object, classified as Sy2/H\,II, and with a bright nucleus. However, it is shown to be resolved (MM07), so no PSF subtraction was attempted. It has a clumpy arc, at 0.5 arcsec southeast of the nucleus, which is very likely a string of star-formation regions. A bright area of diffuse emission is also remarkable, visible in the F330W image up to 1.25 arcsec from the nucleus, in the opposite direction. Although the morphology in the [OIII] band slightly resembles that of F330W image, the ionised gas emission can explain only a small amount of the near-UV flux. With factors as high as 0.9, the bulk of the nuclear and extended emission remains unsubtracted, as well as the star-forming arc. Athough the literature data suggest a scaling factor of 0.36, we find more likely a higher value of about 0.4-0.6. Using {\bf f}=0.5 yields a contribution of the ionised gas of $\sim$28 per cent over the total. 
This galaxy is relatively far, with a radial velocity of 8671~km/s (from NED), which shifts the bright [OIII] lines into the F547M wavelength range. Thus we expect some degree of contamination of the F547M exposure by the cited emission lines. The NICMOS image is dominated by a bright nucleus. As in other cases we perform a PSF subtraction until removing the diffraction patterns and not leaving strong residuals. However, any analysis in the inner $\sim$0.5 arcsec would have strong uncertainties. Due to these two factors, we avoid the nucleus and we focus in studying the extended emission to the northwest. We have performed aperture photometry in two circular apertures of radius 0.25 arcsec at $\sim$0.8 arcsec ($\sim$500\,pc) from the nucleus. The analysis yields very blue colours ({\it U--V}$\sim$-0.18, {\it V--H}$=$-0.93; see Fig.~\ref{fig:result}) compatible with the light being dominated by a blue component (up to 90 per cent of the near-UV flux, once subtracted the ionised gas contribution). For the measurements we used the F330W after subtracting the [OIII] image with an scaling {\bf f}=0.50. We have checked that a change of 0.1 in {\bf f} leads to a change of 0.05~magnitudes in the U--V colour, so the result is quite robust against the assumed {\bf f}, while it constitutes a lower limit of the importance of the blue component due to the possible contamination of the F547M image discussed above. There is obvious star-formation around this Seyfert nucleus, as evidenced by the clumpy arc to the southeast and its spectroscopic classification. Previous studies have shown an important contribution of young stellar population in the inner 2 arcsec of this object (Cid Fernanades et al.~2001; Gonz\'alez Delgado et al.~2001). However, taking into account the polarimetry data in the bibliography for this object (Section~\ref{sec:scat}), and the lack of features in the extended emission to the northwest, we speculate that this last emission is scattered light from the AGN, which explains its very blue U--V colour. Actually, the colours are those of an almost pure power-law with index a=1 (see Fig.~\ref{fig:degPL9Ma}). Another possibility would be the existence of a very young stellar population, although there is no sign of clumps, knots or point-like sources, as it is common in star-forming environments. The young stellar population may come from disrupted star clusters in the nuclear environment, although the presence of knots in the southeast arc indicates that young star-clusters do not necessarily get disrupted at that distance ($\sim$100--150\,pc) from the nucleus. We do not discard that the contribution of young stars is larger in the inner 200\,pc, which would increase the relative contribution of star formation in studies at lower resolution.

\subsection{UGC\,1214 (Sy2)} 
We used the procedure explained above, with {\bf f}=0.30, and end up with a residual image free from the filamentary structure observed in both [OIII] and F330W images. This is the best example in which a reasonable scaling of the [OIII] image can account for all the filamentary structure. From this we deduce that $\sim$40 per cent of the near-UV light comes from nebular continuum and [NeV] emission. 
The resulting image shows a central oval structure (see Fig.~\ref{fig:all3_01}) of diffuse light. Although this object has no F547M exposure, it does have a NICMOS image which is dominated by the bulge. As in other cases, we reduced the NICMOS image and subtracted the central point-like source, ending up with a smooth bulge. Although we cannot have both {\it U--V} and {\it V--H} colours for this object, we verified, using SSP models, that the colour F330W--F160W ({\it U--H}) behave monotonically with age.
 We performed aperture photometry and were able to trace a colour profile up to 2.5 arcsec ($\sim$1200~pc). From 0.5 to 2.5 arcsec the profile stays fairly flat around \mbox{{\it U--H}=0.9}, and in the inner 0.5 arcsec there is a blue drop of 1.5 magnitudes. This implies that the extended emission can be attributed to the underlying bulge, and that there exists a compact blue region in the very centre of the galaxy. A simple proof with synthetic photometry similar to the one used in Fig.~\ref{fig:degPL9Ma}, but only with one colour, shows that, in order to have a drop of 1.5 magnitudes in the centre, it is necessary a contribution of $\sim$70 per cent from a young population or a power-law. On the other hand, Cid Fernandes et al.~(2001) studied the stellar population in the inner 2 arcsec, reporting a contribution of young/intermediate age stellar population and featureless continuum of 10 per cent of the light at 486.1\,nm. Therefore, the previously reported young stellar population in this object may be located in the inner 0.5 arcsec.

\subsection{UGC\,6100 (Sy2)} 
A range of {\bf f} between 0.2 and 0.3 removes what seems to be the ionised gas emission. A higher fraction starts to leave patchy negative residuals near the nucleus. There is an important nuclear contribution which can be attributed to a mechanism different from the NC or [NeV] line emission, unless there is a strong variation in the value of {\bf f} in these small scales, due to a varying ionisation parameter and/or thick obscuring patches. 
The photometric analysis suggests that the remaining light may probably be explained by stellar emission. As can be appreciated in the colour profiles of Fig.~\ref{fig:U-V}, there is a constant {\it U--V} colour gradient of \mbox{$\sim$~-0.12\,mag/100\,pc} extending out more than 1.5 kpc from the nucleus. However, in the inner 100\,pc we find a small red cusp, that is wider than the PSF FWHM. In {\it V--H} colour the behaviour is the opposite, with a slight positive gradient ($\sim$~0.03\,mag/100\,pc), and a central blue dip. This galaxy is listed in NED as of possible Sa type, so the {\it U--V} gradient is to be expected. It is unlikely that the colours of the inner 100\,pc are caused by an excessive subtraction of the UV emission, due to the small {\bf f} used (0.25). Moreover, the [OIII] image is smoother in the centre, and the feature remains when using a smaller {\bf f}. Instead, we suspect that the central colours may be due to contamination of the F547M image by emission lines. This is because the redshift is high enough to make the strong [OIII]$\lambda$5007 fall in the blue wing of the filter, in a position where it has almost 70 per cent of its peak transmission. However, the results beyond 100\,pc from the nucleus are quite robust, and there is no need for other mechanism to be invoked, apart from the stellar emission of bulge and disc components.

\subsection{UGC\,12138 (Sy1.8)} 
The emission is quite compact, concentrated in less than 1 arcsec. The reasonable values of {\bf f} are too small to eliminate the nuclear emission in this case. Forcing a high value of {\bf f} ($>0.8$) the nuclear emission is subtracted, although the extended emission around it remains, and also negative residuals start to appear with {\bf f}$\geq$0.5. Using a scaling factor of 0.35 yields that the contribution of the ionised gas is $\sim$18 per cent of the total near-UV light (once the Sy nucleus is subtracted). So most of the emission is not due to ionised gas in this object. This remaining light constitutes a compact halo around the nucleus, which is visible in the two bluer bands (F330W and F547M). For this object it is worthwhile to perform a PSF subtraction in all the three bands, in order to study the halo close to the nucleus. Then, we performed circular aperture photometry in an annulus around the nucleus, with inner and outer radii of 0.35 arcsec and 0.75 arcsec, that correspond to 175~pc and 375~pc, respectively. The inner radius is large enough to avoid uncertainties due to the subtraction of the nuclear contribution in the bluer bands. The major uncertainties will come by the adopted scaling value for the [OIII] image. Altough in this object the estimate of the ionised gas contribution has a big uncertainty, we will use a factor {\bf f}=0.35, that seems to work well for the objects with ionisation cones. The colours of the extended emission ({\it U--V}$=$1, {\it V--H}$=$-0.98) are similar to those in NGC\,5347, which has also a similar morphology. With the caveat discussed above, we interpret the colour as due to an old bulge with a contribution of scattered nuclear light.


 \begin{center}
 \begin{landscape}
 \begin{table}
 \centering
  \begin{minipage}{230mm}
   \caption{Summary of the results. }
   \begin{tabular}[c]{@{}lccccccccccp{37mm}p{37mm}@{}}
   \hline
 Object    & Seyfert & Scale & E(B-V) & Nucleus & Nuclear  & m(1kpc) & likely f & $\Delta$f & Fneb/Ftot & $\Delta$Fn/Ft & Polarimetric studies & Photometry of  \\
 name      &  type   & ("/kpc) &      &         & contrib. & (mag)   &          &             &         &               &                      & the residuals \\
   \hline
IC\,5063   &  Sy2    & 4.61  & 0.061  &  No   &      &  15.66  &   0.25   & 0.05 &   0.23   &  0.05  & P$\sim$1-2\% in the optical$^d$ & Remaining light compatible with bulge population  \\
Mrk\,6     &  Sy1.5  & 2.55  & 0.136  &  Yes  & 0.72 &  15.94  &   0.20   & 0.05 &   0.62   &  0.16  & P$\sim$1\% at red wavelengths$^e$ &  \\
Mrk\,915   &  Sy1    & 2.08  & 0.063  &  Yes  & 0.72 &  16.22  &   0.30   & 0.05 &   0.30   &  0.05  &         &  \\
NGC\,1320  &  Sy2    & 5.85  & 0.047  &  No   &      &  14.80  &          &      & $<$ 0.15$^c$  &   &         & Old bulge population with a significant bluer component, likely star-formation.  \\
NGC\,3393  &  Sy2    & 4.10  & 0.075  &  No   &      &  15.00  &   0.35   & 0.05 &   0.33   &  0.10  &         &    \\
NGC\,3516  &  Sy1.5  & 5.05  & 0.042  &  Yes  & 0.60 &  14.46  &          &      & $<$ 0.11$^c$  &   &         & Compact diffuse emission with bulge colours.   \\
NGC\,4253  &  Sy1.5  & 3.66  & 0.020  &  Yes  & 0.32 &  15.24  & 0.41$^a$ &      & $<$ 0.16$^c$  &   &         &     \\
NGC\,4593  &  Sy1    & 5.32  & 0.025  &  Yes  & 0.71 &  14.53  &          &      & $<$ 0.08$^c$  &   & P$<$1\% at red wavelengths$^e$ &   \\
NGC\,5347  &  Sy2    & 5.62  & 0.021  &  No   &      &  16.26  & 0.40$^b$ & 0.05 & $<$ 0.08$^c$  &   &         & Old population plus a 10-30\% contribution of bluer component. Indicates reddening effects or scattered AGN light.     \\
NGC\,5548  &  Sy1.5  & 2.79  & 0.020  &  Yes  & 0.74 &  15.70  &          &      & $<$ 0.25$^c$  &   & P$<$1\% at red wavelengths$^e$ &    \\
NGC\,7212  &  Sy2    & 1.86  & 0.072  &  No   &      &  16.61  &   0.20   & 0.05 &  0.53  &  0.13  & P$\sim$10\%. Scattered light may contribute $\lesssim$30\% $^f$ & Clear star-formation. Scattered light may be present.  \\
NGC\,7674  &  Sy2    & 1.72  & 0.059  &  No   &      &  15.58  &   0.50   &  0.15 &  0.28  &  0.08  & P$\gtrsim$10\%. Scattered light may contribute $\lesssim$40\% $^f$ & Very blue residuals. Both, star-formation and scattered light, are present.   \\
UGC\,1214  &  Sy2    & 2.95  & 0.023  &  No   &      &  15.59  &   0.30   &  0.05 &  0.39  &  0.07  &         & Old bulge plus a blue contribution in the inner 0.5''.   \\
UGC\,6100  &  Sy2    & 1.66  & 0.012  &  No   &      &  17.04  &   0.25   &  0.05 &  0.18  &  0.04  &         & Mostly stellar emission.   \\
UGC\,12138 &  Sy1.8  & 1.99  & 0.085  &  Yes  & 0.72 &  16.13  &   0.35   &  0.10 &  0.18  &  0.05  &         & Mostly bulge emission with a small bluer contribution. \\
 \hline
 \label{tab:res}
 \end{tabular}

Col.(1)\,: galaxy name; Col.\,(2): nuclear activity type; Col.\,(3): inverse scale, as angular size of the studied region, calculated from Table\,\ref{tab:oiii_sample}; Col.\,(4): galactic reddening, E(B-V), from NED; Col.\,(5): occurrence of a point-like nucleus in the near-UV; Col.\,(6): contribution of the nucleus to the total near-UV light inside 1\,kpc; Col.\,(7): F330W magnitude of the inner 1\,kpc (after nuclear subtraction), corrected from galactic extinction; Col.\ (8): estimated value for the scaling factor {\bf f}; Col.\,(9): estimated uncertainty of scaling factor {\bf f}; Col.\,(10): calculated contribution of nebular emission to the total emission (after nuclear subtraction) in the inner 1\,kpc; Col.\,(11): uncertainty of (10), calculated from (9) and considering an uncertainty of 5 per cent of the peak value when scaling the PSF for the nuclear subtraction.; Col.\,(12): polarimetric studies for some objects of the sample; Col.\,(13): summary of the results from the photometry of the images after subtraction of the ionised gas component.
\medskip

 $^a$ Literature based value.\\
 $^b$ Calculated for the arc of emission to the north-east.\\
 $^c$ Upper limits are normally calculated for f=0.6, either in the inner kpc or in the region in which we have enough [OIII] signal-to-noise.\\
 $^d$ Reference: Inglis et al.~(1993)\\
 $^e$ Reference: Smith et al.~(2002)\\
 $^f$ Reference: Tran (1995)\\

\end{minipage}
 \end{table}
\end{landscape}
\end{center}


\section{SUMMARY AND DISCUSSION}
\label{sec:summary}

The near-ultraviolet morphology of the 15 objects studied in this work has been described in Mu\~noz Mar\'{i}n et al.~(2007). This work is complementary to it, as we intend to gain insight into the light emitting processes that generate the particular morphology of each object. 
In the nuclei of Seyfert type 1 to 1.5, the dominant component is the AGN emission, visible as a bright point-like source. In these objects the nucleus has to be removed prior to any further analysis.

In order to understand the diffuse light emission we start by focussing on the ionised gas emission. We have used the narrow band [OIII] images from Schmitt et al.~(2003) to remove this component. The [OIII] images were scaled in order to model and subtract the predicted contribution from the [NeV] lines and nebular continuum from the F330W image. An estimation of a reasonable range of scaling factors was presented in Section~\ref{sec:NC}.  A comparison of the narrow band images (tracing the ionised gas), with the near-UV broad band images, show a strong similarity in many cases. We have demonstrated, by scaling accordingly the [OIII] images, that the emission of the ionised gas can explain the extended emission visible in several objects. This is the case of the filamentary structure in IC\,5063, NGC\,3393 or UGC\,1214 (Mrk\,573), as well as parts of the more diffuse emission in Mrk\,6, Mrk\,915, NGC\,5347 and NGC\,7212. These regions of ionised gas are large, extending from 100\,pc to 1\,kpc from the nucleus, and in general can be accounted for with a scaling factor for the [OIII] image (in the range f$\sim$0.2--0.4), consistent with the expected level of line emission and nebular continuum contamination.

\begin{figure}
\includegraphics[angle=0,width=0.5\textwidth]{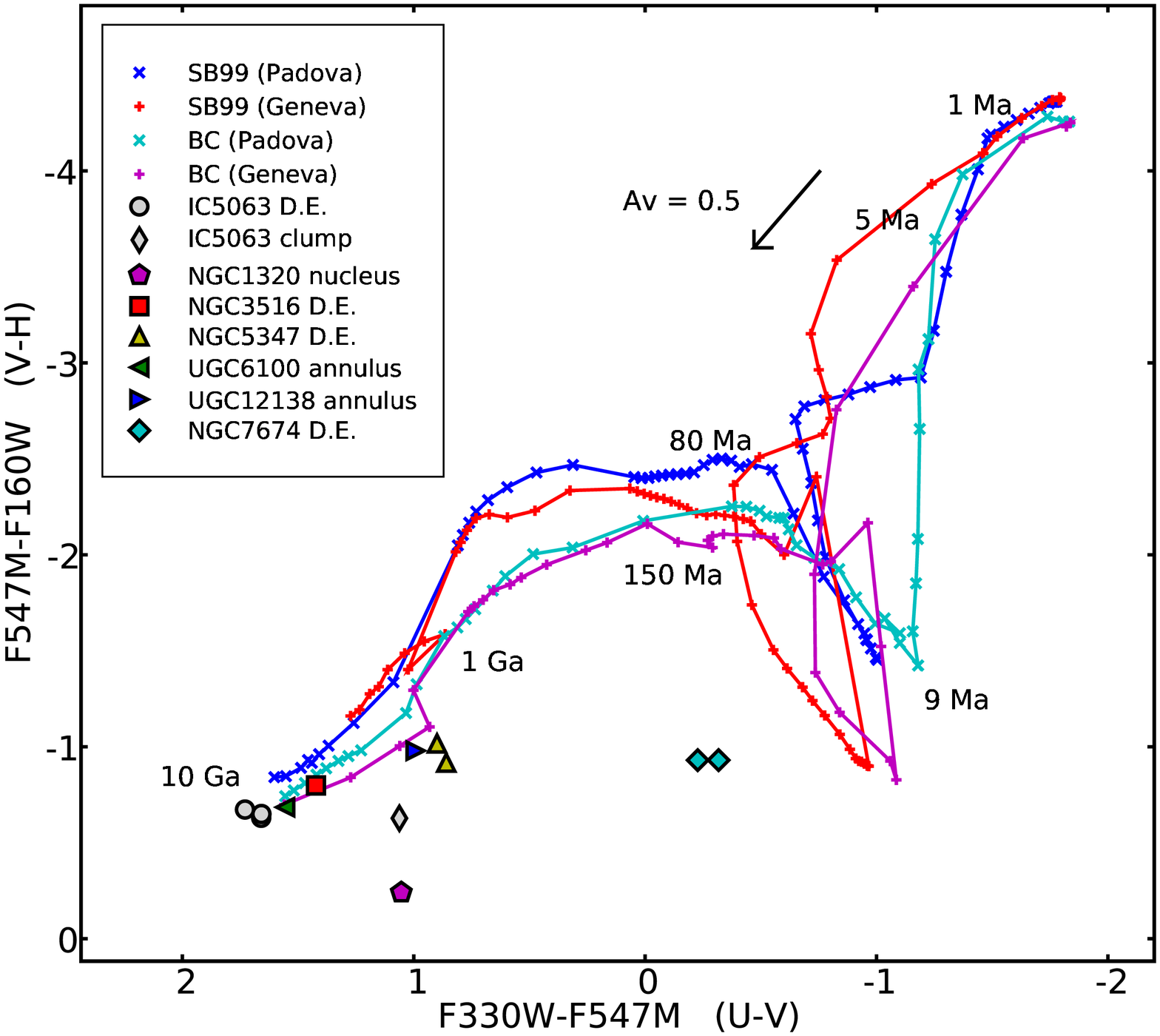}
\caption{Same colour-colour diagram as Fig.~\ref{fig:UV-VH_comp}, but adding the photometric results for several objects. Different colours and symbols stand for different objects and regions, which are labelled in the plot legend, in which `D.E.' stands for diffuse emission. The photometric points are corrected for galactic extinction with the law of Cardelli et al.~(1989). The reddening vector corresponding to A$_V$=0.5 is also plotted for reference.}
\label{fig:result}
\end{figure}

When possible, we have performed a photometric study in three bands: near-UV (ACS F330W), optical (WFPC2 F547M), and near-IR (NICMOS F160W). The results are summarised in a colour-colour diagram, in Fig.~\ref{fig:result}. We have measured colours around F160W--F547M$\sim$-1 and F547M--F330W$\sim$0.8--1.8, in STMAG, for several objects. These colours are consistent with a several Ga old stellar population. We interpret this as an indication that the diffuse emission remaining after estimating and subtracting the ionised gas contribution, can be attributed to the underlying stellar population of the bulge. This is the case of objects such as IC\,5063, NGC\,1320, NGC\,3516, UGC\,1214, UGC\,6100, and UGC\,12138. 

In a number of objects we also find a strong near-UV emission extending less than 100\,pc from the nucleus (see for example NGC\,4593). These are mostly Sy1 in which even after the best nuclear-PSF subtraction, there remains a compact but resolved emission. We have already argued that we do not think this is an artifact of the PSF subtraction of the nucleus, but a real feature. However, it is true that the nuclear subtraction generates uncertainties that preclude any accurate analysis of the very inner region. In these cases, we cannot say much about the origin of this light. We have made some calculations to set an upper limit to the contribution of the ionised gas emission, as indication for a different mechanism. A strong change in the ionisation parameter (higher towards the centre) has not been ruled out, though.

The physical regions studied in this work are located between 100\,pc and 1\,kpc from the Seyfert nuclei. At these scales, the scattered light from the AGN does not seem to contribute in a significant fraction to the total light, in the majority of the cases. We demonstrated that, in general, an ionised gas contribution is very difficult to distinguish from a young stellar population with a photometric analysis and the filters used.
However, we suspect that in the case of NGC\,7674, the scattered light from the AGN is responsible for the near-UV emission to the northwest of the nucleus. This is the only evidence for scattered AGN light that we find among the objects studied, although absence of evidence is different from evidence of absence, and scattered light could play an important role in other objects with an unidentified blue component, as for example NGC\,7212. 
This result differs from those in radio galaxies studies. For example, Tadhunter et al.~(2002) find that the objects with largest UV polarisation tend to have the largest emission line luminosities. In Cygnus A, Ogle et al.~(1997), find reflection nebulae in polarised light that have a similar scale and general morphology to the ionisation cone seen in emission lines. There is a possible way out of this. Dust is expected to correlate with the gas distribution, and therefore some scattered emission might have been subtracted with the ionised gas. That would explain high values of the scaling factor for the [OIII] image, as in the case of NGC\,1320, and at least in part, the positive residuals of some of the objects. However, as discussed above, ionised gas emission alone can explain the cases in which the near-UV morphology matches the [OIII] emission.

 Regarding a possible starburst component, we have not found, in general, unequivocal evidence of extended unresolved young stellar populations. However there are several objects, such as IC\,5063, NGC\,1320, NGC\,3393, NGC\,5347, and UGC\,1214, which show spectroscopic evidence for young or intermediate age stellar population (Cid Fernandes et al.~2001; Cid Fernandes et al.~2004; Gonz\'alez Delgado \& P\'erez 1996b) but do not show obvious star clusters in the near-UV images. That would be the case of a population of small star clusters appearing unresolved in our images, or a very high rate of cluster disruption in the nuclear region. 
Star-forming regions and massive star clusters stand out as compact knots in the near-UV image. They make a significant contribution in a fraction of objects (4/15, or $\sim$27 per cent); namely, NGC\,4253, NGC\,5548, NGC\,7212, and NGC\,7674. In addition there are 2 objects (NGC\,4593 and UGC\,6100) in which star clusters are identified but lay at kpc distance from the centre of the galaxy, far away of the region studied and thus they do not contribute to the near-UV emission in the nuclear region. This is why we leave out these two objects in the general conclusions. Therefore $\sim$40 per cent of the sample (6/15 objects, including the two objects with clusters further than 1\,Kpc) show circumnuclear star clusters, while in the original sample we find star clusters in around 70 per cent of the sample. This difference can be attributed to a bias due to the fact that the sample analysed here is only a small subsample of the larger sample, selected on the basis of availability of [OIII] data.

There exists the possibility that a young stellar population is completely diluted in the bulge light and we cannot detect it. 
Emission from the ionised gas, as well as the underlying bulge stellar population are enough to explain the diffuse light in most cases. On the contrary, the clumps and compact emission are likely due to star-formation or AGN light. 

The results in Table~\ref{tab:res} show no apparent trend between the Seyfert type and the magnitudes calculated, such as estimated scaling factor for the [OIII] image, the ionised gas contribution to the total near-UV light, or the flux in the inner kpc. However, we have to take into account the small size of the sample and the difficulties to study some Sy1 for which we could set only upper limits to the ionised gas contribution.


\section{CONCLUSIONS}
\label{sec:conclusions}

Our main conclusions are:\\
- For $\sim$50 per cent of the galaxies of the sample (8/15), emission of ionised gas from the NLR dominates the near-UV extended emission reaching up to 1\,kpc from the nucleus.\\
- For 7 galaxies, we could obtain broad-band photometry of the UV images after the subtraction of the ionised gas emission; for 6 of them, colours of the extended emission left after the subtraction of the gas emission are consistent with those of an old stellar population; among these galaxies there are also 5 objects for which we find evidence of an additional contribution of either scattered nuclear light or a diffuse young stellar population.\\
- Only for 1 galaxy we have found that scattered light dominates the extended emission, although this kind of emission might be present at a lower level in a large fraction of the objects.\\
- For 4/15 galaxies of the sample, star forming regions appear as compact knots surrounding the nucleus and contribute significantly to the near-UV light in the inner kpc.

In summary, there is not a unique and dominating mechanism responsible for the near-UV emission. Each object deserves a particular study and a close inspection, as we find the most varied morphologies. A deeper study would require the use of spectroscopy, and optimally, the use of IFU data with enough resolution and polarisation observations.

\section*{ACKNOWLEDGEMENTS}
We gratefully thank Clive Tadhunter and Valentina Luridiana for usefull comments that helped to improve this paper. We would also like to acknowledge the comments of the referee, Makoto Kishimoto, which in particular helped to shape and present the conclusions of this work.
  V. M. M. M.'s research has been funded by the Spanish Research Council (CSIC) under the I3P grant program. This work was supported in part by the Spanish Ministerio de Educaci\'on y Ciencia under grant AYA 2007-64712. 
 This work was carried out in part at the Instituto de F\'{i}sica of the Universidade Federal do Rio Grande do Sul (Porto Alegre), where we enjoyed the Brazilian hospitality and made use of the university's facilities.

\label{lastpage}


\clearpage

\end{document}